\newif\ifextended 
\newif\ifauthor 

\extendedtrue\authortrue

\documentclass[runningheads]{llncs}

\usepackage[utf8]{inputenc}
\usepackage[T1]{fontenc}

\usepackage{graphicx}

\usepackage{mathtools}
\usepackage{amsfonts}
\usepackage{amssymb}

\usepackage[hidelinks]{hyperref}

\usepackage{multicol}

\usepackage{bbding}

\usepackage{listings}
\lstdefinestyle{alg}{%
  basicstyle=\footnotesize\sffamily,
  morekeywords={let,match,if,else,in,then,with,function,for,return,Input,Output},
}
\lstdefinestyle{ppl}{%
  basicstyle=\scriptsize\ttfamily,
  showlines=true,
}
\lstdefinelanguage{CorePPL}{%
  basicstyle=\ttfamily\footnotesize,
  morekeywords={mexpr,let,assume,observe,true,false,include,type,con,in,lam,match,with,then,else,never,recursive,weight,resample,if,infer},
  morecomment=[l]{--},
  commentstyle = \color{olive},
}
\lstdefinelanguage{Anglican}{
  language=Lisp,
  basicstyle=\ttfamily\footnotesize,
  morekeywords={sample,observe,ns,use,gen-class,def,defdist,defm,defn,defquery,if,defquerythrow,with-primitive-procedures},
  commentstyle = \color{olive},
}
\lstdefinelanguage{WebPPL}{
  basicstyle=\ttfamily\footnotesize,
  keywords={typeof, new, true, false, catch, function, return, null, catch, switch, var, if, in, while, do, else, case, break},
  ndkeywords={class, export, boolean, throw, implements, import, this},
  comment=[l]{//},
  morecomment=[s]{/*}{*/},
  commentstyle = \color{olive},
  morestring=[b]',
  morestring=[b]"
}

\usepackage{pgf}

\usepackage{pgfplots}
\pgfplotsset{compat=1.16}
\usepgfplotslibrary{statistics}

\usepackage{algorithm}

\usepackage{subcaption}

\newcommand\doubleplus{+\kern-0.7ex+}
\newcommand{\ttt}[1]{\texttt{\textup{#1}}}
\newcommand{\tsf}{\textsf}
\newcommand{\tsc}{\textsc}
\newcommand{\tbf}{\textbf}
\newcommand{\mi}{\mathit}
\newcommand{\term}{\textbf{\upshape\tsf{t}}}
\newcommand{\termv}{\textbf{\upshape\tsf{v}}}
\newcommand{\termanf}{\term_\textrm{ANF}}
\newcommand{\termanfty}{T_\textrm{ANF}}
\newcommand{\termid}{\term_\textrm{id}}
\newcommand{\absval}{\textbf{\tsf{a}}}
\newcommand{\cstr}{\textbf{\tsf{c}}}
\newcommand{\suspend}{\mi{suspend}}
\newcommand{\false}{\textrm{false}}
\newcommand{\true}{\textrm{true}}
\newcommand{\s}{\enspace}

\newcommand\texample{\term_\textrm{example}}
\newcommand\varsexample{\tsf{vars}_\textrm{example}}

\newcommand\Susassume{\textrm{Suspension}_\ttt{assume}}
\newcommand\Susweight{\textrm{Suspension}_\ttt{weight}}

\newcommand\sem[3]{{\displaystyle\hspace{1mm}\prescript{#2\vphantom{#3}}{\vphantom{#1}}{\Downarrow}^{#3\vphantom{#2}}_{#1}\hspace{1mm}}}
\newcommand\concat{\mathbin\Vert}

\usepackage[absolute]{textpos}

\begin{document}

\title{Suspension Analysis and Selective Continuation-Passing Style for Universal Probabilistic Programming Languages}
\titlerunning{Suspension Analysis and Selective CPS for Universal PPLs}

\makeatletter
\def\orcidID#1{\href{http://orcid.org/#1}{\protect\raisebox{-1.25pt}{\protect\includegraphics{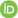}}}}
\makeatother

\author{%
  Daniel~Lundén\inst{1}(\raisebox{-2.4pt}{\Envelope})\orcidID{0000-0003-3127-5640} \and %
  Lars~Hummelgren\inst{2} \and %
  Jan~Kudlicka\inst{3}\orcidID{0000-0003-3806-4950} \and %
  Oscar~Eriksson\inst{2}\orcidID{0009-0001-0678-549X} \and %
  David~Broman\inst{2,4}\orcidID{0000-0001-8457-4105}
}

\authorrunning{D. Lundén et al.}

\institute{%
  Oracle, Stockholm, Sweden, \email{daniel.lunden@oracle.com} \and
  EECS and Digital Futures, KTH Royal Institute of Technology, Stockholm, Sweden, \email{\{larshum,oerikss,dbro\}@kth.se} \and
  Department of Data Science and Analytics, BI Norwegian Business School, Oslo, Norway, \email{jan.kudlicka@bi.no} \and
  Computer Science Department, Stanford University, California, USA \email{broman@stanford.edu}
}

\maketitle

\ifauthor
\begin{textblock*}{1.2\textwidth}(1in + \oddsidemargin - 0.1\textwidth,10pt)
  \noindent
  \scriptsize
  This is an author-prepared version\ifextended, extended with appendices and minor additions to the main text\fi.
  © The Author(s) 2024.
  This version of the contribution\ifextended, except the extensions,\fi\ has been accepted for publication at ESOP 2024, after peer review, but is not the Version of Record (the version published by Springer) and does not reflect post-acceptance improvements, or any corrections.
  The Version of Record is available online at: \url{https://doi.org/10.1007/978-3-031-57267-8_12}.
\end{textblock*}
\fi

\begin{abstract}
  Universal probabilistic programming languages (PPLs) make it relatively easy to encode and automatically solve statistical inference problems.
  To solve inference problems, PPL implementations often apply Monte Carlo inference algorithms that rely on execution suspension.
  State-of-the-art solutions enable execution suspension either through (i) continuation-passing style (CPS) transformations or (ii) efficient, but comparatively complex, low-level solutions that are often not available in high-level languages.
  CPS transformations introduce overhead due to unnecessary closure allocations---a problem the PPL community has generally overlooked.
  To reduce overhead, we develop a new efficient selective CPS approach for PPLs.
  Specifically, we design a novel static suspension analysis technique that determines parts of programs that require suspension, given a particular inference algorithm.
  The analysis allows selectively CPS transforming the program only where necessary.
  We formally prove the correctness of the analysis and implement the analysis and transformation in the Miking CorePPL compiler.
  We evaluate the implementation for a large number of Monte Carlo inference algorithms on real-world models from phylogenetics, epidemiology, and topic modeling.
  The evaluation results demonstrate significant improvements across all models and inference algorithms.
\end{abstract}


\section{Introduction}\label{sec:intro}

\noindent
Probabilistic programming languages (PPLs), such as Anglican~\cite{wood2014new}, Birch~\cite{murray2018automated}, WebPPL~\cite{goodman2014design}, Stan~\cite{carpenter2017stan}, Pyro~\cite{bingham2019pyro}, and Gen~\cite{towner2019gen}, make it possible to encode and solve statistical inference problems.
Such inference problems are of significant interest in many research fields, including phylogenetics~\cite{ronquist2021universal}, computer vision~\cite{kulkarni2015picture}, topic modeling~\cite{blei2003latent}, inverse graphics~\cite{gothoskar20213dp3}, and cognitive science~\cite{goodman2016probabilistic}.
A particularly appealing feature of PPLs is the separation between the inference problem specification (the language) and the inference algorithm used to solve the problem (the language implementation).
This separation allows PPL users to focus solely on encoding their inference problems while inference algorithm experts deal with the intricacies of inference implementation.

Implementations of PPLs apply many different inference algorithms.
Monte Carlo inference algorithms---such as Markov chain Monte Carlo (MCMC)~\cite{gilks1995markov} and sequential Monte Carlo (SMC)~\cite{doucet2001sequential}---are popular due to their asymptotic correctness and relative ease of implementation for \emph{universal}\footnote{%
  A term that first appeared in Goodman et al.~\cite{goodman2008church}, indicating expressive PPLs where the number and types of random variables are not always known statically.%
} PPLs.
The central idea behind all Monte Carlo methods in PPLs is to execute probabilistic programs multiple times to generate \emph{samples} that approximate the target distribution for the encoded inference problem.
However, repeated execution is expensive, and PPL implementations must avoid unnecessary overhead.


Monte Carlo algorithms often need to \emph{suspend} executions.
For example, MCMC algorithms can suspend at \emph{random draws} in the program to avoid unnecessary re-execution when proposing new executions, and SMC algorithms can suspend at \emph{likelihood updates} to \emph{resample} executions.
Languages such as WebPPL~\cite{goodman2014design} and Anglican~\cite{wood2014new}, and the approach described by Ritchie et al.~\cite{ritchie2016c3}, apply \emph{continuation-passing style (CPS)} transformations~\cite{appel1991compiling} to enable arbitrary suspension during execution.
The main benefit of CPS transformations is that they are relatively easy to implement in functional programming languages.
However, one disadvantage with CPS transformations is that high-performance low-level languages, without higher-order functions, do not support them.
For this reason, there are also more direct low-level alternatives to CPS, including non-preemptive multitasking (e.g., coroutines~\cite{ge2018turing}) and PPL control-flow graphs~\cite{lunden2022compiling}.
These more direct alternatives can additionally avoid much of the overhead resulting from CPS\footnote{%
  Note that CPS only results in overhead if programs reify the continuations at runtime to, e.g., suspend computations.
  Traditional CPS-based compilers often only use CPS as an intermediate form during compilation, which does not result in runtime overhead.
}, but are more complex to implement.

We consider how to bridge the performance gap between CPS-based PPLs and lower-level PPLs that rely on, e.g., direct implementation of coroutines.
We consider optimizations at the CPS transformation level, and not the translation from CPS-based PPLs to lower-level representations.
CPS overhead is a result of closure allocations for continuations.
We make the important observation that PPLs do not require the arbitrary suspensions provided by full CPS transformations.
Most Monte Carlo inference algorithms require suspension only in very specific parts of programs.
Current state-of-the-art CPS-based PPLs do not consider inference-specific suspension requirements to reduce CPS overhead.

We design a new static suspension analysis and a new selective CPS transformation for PPLs that together significantly reduce runtime overhead compared to a traditional full CPS transformation.
Current state-of-the-art functional PPLs that use CPS for execution suspension can therefore greatly benefit from our new approach.
The suspension analysis identifies all parts of programs that may require suspension as a result of applying a particular inference algorithm.
We formalize the suspension analysis algorithm using a core PPL calculus equipped with a big-step operational semantics.
Specifically, the challenge lies in capturing how suspension requirements propagate through the program in the presence of higher-order functions.
Furthermore, we formalize the selective CPS transformation and justify its correctness when guided by the suspension analysis.
Prior work on selective CPS for general-purpose programming languages, e.g., by Nielsen~\cite{nielsen2001selective} and Asai and Uehara~\cite{asai2017selective}, focuses on analyses based on type systems and type inference.
In contrast, we instead build our suspension analysis using 0-CFA~\cite{shivers1991control} and it operates directly on an untyped calculus.

Overall, we (i) prove that the suspension analysis is correct, (ii) show that the resulting selective CPS transformation gives significant performance gains compared to using a full CPS transformation, and (iii) show that the overall approach is directly applicable to a large set of inference algorithms.
Specifically, we evaluate the approach for the following inference algorithms: likelihood weighting, the SMC bootstrap particle filter, the SMC alive particle filter~\cite{kudlicka2019probabilistic}, aligned lightweight MCMC~\cite{lunden2023automatic,wingate2011lightweight}, and particle-independent Metropolis--Hastings~\cite{paige2014compilation}.
We consider each inference algorithm for four real-world models from phylogenetics, epidemiology, and topic modeling.

We implement the suspension analysis and selective CPS transformation in Miking CorePPL~\cite{lunden2022compiling,broman2019vision}.
Similarly to WebPPL and Anglican, the implementation supports the co-existence of many inference problems and applications of inference algorithms to these problems within the same program.
However, compared to full CPS, such programs are more challenging to handle with selective CPS, as the CPS transformation of an inference problem also depends on the applied inference algorithm---different inference algorithms generally require different suspensions.
To complicate things further, different inference problems may share some code, or the PPL user may apply two different inference algorithms to the same inference problem.
The compiler must then apply different CPS transformations to different parts of the program, and sometimes even many different CPS transformations to separate copies of the \emph{same} part of the program.
To solve this, we develop an approach that, for any given Miking CorePPL program, \emph{extracts} all possible inference problems and corresponding inference algorithm applications.
This extraction procedure allows the correct application of selective CPS throughout the program.

In summary, we make the following contributions.
\begin{itemize}
  \item
    We design, formalize, and prove the correctness of a suspension analysis for PPLs, where the suspension requirements come from a given inference algorithm (Section~\ref{sec:sus}).
  \item
    We design and formalize a new selective CPS transformation for PPLs.
    Compared to full CPS, selectively CPS transforming PPL programs guided by the suspension analysis significantly reduces runtime overhead resulting from unnecessary closure allocations (Section~\ref{sec:cps}).
  \item
    We implement the suspension analysis and selective CPS transformation in the Miking CorePPL compiler.
    Unlike full CPS, selective CPS introduces challenges for probabilistic programs containing many inference problems and inference algorithm applications.
    We implement an approach that correctly applies selective CPS to such programs by extracting individual inference problems (Section~\ref{sec:implementation}).
\end{itemize}
Section~\ref{sec:evaluation} presents the evaluation and its results for the implementations in Miking CorePPL, Section~\ref{sec:related} discusses related work in more detail, and Section~\ref{sec:conclusion} concludes.
We first consider a motivating example in Section~\ref{sec:motivating} and introduce the underlying PPL calculus in Section~\ref{sec:ppl}.

\pgfplotsset{%
  probplot/.style={%
    ytick=\empty,
    width=50mm,
    height=40mm,
    tick style={draw=none},
    tick label style={font=\scriptsize},
    label style={font=\scriptsize},
    x label style={at={(axis description cs:0.5,-0.3)},anchor=north},
    axis line style={draw=none},
  }
}
\begin{figure}[tbp]
  \lstset{%
    basicstyle=\ttfamily\scriptsize,
    showlines=true,
    framexleftmargin=-2pt,
    xleftmargin=2em,
  }%
  \centering
  \begin{minipage}{0.53\textwidth}
  \begin{subfigure}{\textwidth}
    \centering
    \begin{lstlisting}[style=ppl]
let $a_1$ = assume ($\textrm{Beta}$ $2$ $2$) in$\label{line:base:assume}$
let rec $\mi{iter}$ = $\lambda \mi{obs}.$$\label{line:base:recbegin}$
  if $\mi{null}$ $\mi{obs}$ then $()$ else$\label{line:base:ret}$
    weight ($f_{\textrm{Bernoulli}}$ $a_1$ ($\mi{head} \, \mi{obs})$);$\label{line:base:weight}$
    $\mi{iter}$ ($\mi{tail}$ $\mi{obs}$)
in
$\mi{iter}$ $[\true{}$,$\true{}$,$\false{}$,$\true{}]$;$\label{line:base:recend}$
$a_1$$\label{line:base:return}$
    \end{lstlisting}
    \caption{Program $\texample$.}
    \label{fig:running:base}
  \end{subfigure}
  \end{minipage}
  \begin{minipage}{0.43\textwidth}
    \begin{subfigure}{\textwidth}
      \centering
      \begin{tikzpicture}
        \begin{axis}[%
          axis x line*=bottom,
          y axis line style={draw=none},
          width=0.8\textwidth,
          height=2.7cm,
          ymin=0, ymax=2.5,
          xmin=0, xmax=1,
          tick style={draw=none},
          xtick={0,0.5,1},
          ytick=\empty,
          ]
          \newcommand{\xalphap}{2}
          \newcommand{\xbetap}{2}
          \addplot [
            domain=0:1
            ] {%
              x^(\xalphap-1) * (1-x)^(\xbetap-1) /
              ((\xalphap-1)!*(\xbetap-1)!/(\xalphap+\xbetap-1)!)
            };
        \end{axis}
      \end{tikzpicture}
      \caption{Beta(2,2).}
      \label{fig:running:beta}
    \end{subfigure} \\[2mm]
    \begin{subfigure}{\textwidth}
      \centering
      \begin{tikzpicture}
        \begin{axis}[%
          axis x line*=bottom,
          y axis line style={draw=none},
          width=0.8\textwidth,
          height=2.7cm,
          ymin=0, ymax=2.5,
          xmin=0, xmax=1,
          tick style={draw=none},
          xtick={0,0.5,1},
          ytick=\empty,
          ]
          \newcommand{\xalphafour}{5}
          \newcommand{\xbetafour}{3}
          \addplot [
            fill=gray!70,
            domain=0:1
            ] {%
              x^(\xalphafour-1) * (1-x)^(\xbetafour-1) /
              ((\xalphafour-1)!*(\xbetafour-1)!/(\xalphafour+\xbetafour-1)!)
            };
        \end{axis}
      \end{tikzpicture}
      \caption{Distribution of $\texample$.}
      \label{fig:running:posterior}
    \end{subfigure}
  \end{minipage} \\
  \begin{minipage}{0.44\textwidth}
  \begin{subfigure}{\textwidth}
      \begin{lstlisting}[style=ppl]
$\Susassume(\textrm{Beta}$ $2$ $2$, $\lambda a_1$.
  let rec $\mi{iter}$ = $\lambda \mi{obs}.$
    if $\mi{null}$ $\mi{obs}$ then $()$ else
      weight ($f_{\textrm{Bernoulli}(a_1)}$
                ($\mi{head} \, \mi{obs}$));
      $\mi{iter}$ ($\mi{tail}$ $\mi{obs}$)
  in
  $\mi{iter}$ $[\true{}$,$\true{}$,$\false{}$,$\true{}]$;
  $a_1)$
      \end{lstlisting}
      \caption{Suspension at \ttt{assume}.}
      \label{fig:running:assume}
  \end{subfigure}\\
  \begin{subfigure}{\textwidth}
    \begin{lstlisting}[style=ppl]
let $a_1$ = assume ($\textrm{Beta}$ $\!2$ $\!2$) in
let rec $\mi{iter}$ = $\lambda k.$ $\lambda \mi{obs}.$
  if $\mi{null}$ $\mi{obs}$ then $k$ $()$$\label{line:weight:ret}$
  else
    $\Susweight($$\label{line:weight:sus}$
      $f_{\textrm{Bernoulli}(a_1)}$ ($\mi{head} \, \mi{obs}$),
      ($\lambda \_.$ $\mi{iter}$ $k$ ($\mi{tail}$ $\mi{obs}$))$)$$\label{line:weight:k}$
in
$\mi{iter}$ ($\lambda \_.$ $a_1$)
  $[\true{}$,$\true{}$,$\false{}$,$\true{}]$;
    \end{lstlisting}
    \caption{Suspension at \ttt{weight}.}
      \label{fig:running:weight}
  \end{subfigure}
  \end{minipage}
  \begin{minipage}{0.55\textwidth}
  \begin{subfigure}{\textwidth}
    \centering
    \begin{tabular}{c}
      \begin{lstlisting}[style=ppl]
let $k_7$ = $\lambda t_6.$
 let $k_8$ = $\lambda t_7.$
  $\Susassume($$t_7$, $\lambda a_1.$$\label{line:full:susassume}$
   let rec $\mi{iter}$ = $\lambda k_1.$ $\lambda \mi{obs}.$
    let $k_2$ = $\lambda t_1.$
     if $t_1$ then $k_1$ $()$ else
      let $k_3$ = $\lambda t_2.$
       let $k_4$ = $\lambda t_3.$
        let $k_5$ = $\lambda t_4.$
         $\Susweight($$t_4$, $\lambda \_.$$\label{line:full:susweight}$
          let $\!k_6\!$ = $\!\lambda t_5.\!$ $\!\mi{iter}$ $\!k_1\!$ $t_5$ in
          $\mi{tail}_\textrm{CPS}$ $k_6$ $\mi{obs})$
        in $t_2$ $k_5$ $t_3$
       in $\mi{head}_\textrm{CPS}$ $k_4$ $\mi{obs}$
      in ${f_\textrm{Bernoulli}}_\textrm{CPS}$ $k_3$ $a_1$
    in $\mi{null}_\textrm{CPS}$ $k_2$ $\mi{obs}$
   in $\mi{iter}$ ($\lambda \_.$ $a_1$)
        $[\true{}$,$\true{}$,$\false{}$,$\true{}])$
 in $t_6$ $k_8$ $2$
in $\textrm{Beta}_\textrm{CPS}$ $k_7$ $2$
      \end{lstlisting}
    \end{tabular}
    \caption{Full CPS.}
    \label{fig:running:full}
  \end{subfigure}
  \end{minipage}
  \caption{%
    A probabilistic program $\texample$ modeling the bias of a coin.
    Fig.~(a) gives the program.
    The function $f_\textrm{Bernoulli}$ is the probability mass function of the Bernoulli distribution.
    Fig.~(b) illustrates the distribution for $a_1$ at line~\ref{line:base:assume} in (a).
    Fig.~(c) shows the set of (weighted) samples resulting from conceptually running $\texample$ infinitely many times.
    Fig.~(d) and Fig.~(e) show the selective CPS transformations required for suspension at \ttt{assume} and \ttt{weight}, respectively.
    Fig.~(f) gives $\texample$ in full CPS, with suspensions at \ttt{assume} and \ttt{weight}.
    The $_\textrm{CPS}$ subscript indicates CPS-versions of intrinsic functions such as $\mi{head}$ and $\mi{tail}$.
  }
  \label{fig:running}
\end{figure}
\section{A Motivating Example}\label{sec:motivating}
This section introduces the running example in Fig.~\ref{fig:running} and uses it to present the basic idea behind PPLs and how inference algorithms such as SMC and MCMC make use of CPS to suspend executions.
Most importantly, we illustrate the motivation and key ideas behind selective CPS for PPLs.

Consider the probabilistic program in Fig.~\ref{fig:running:base}, written in a functional-style PPL.
The program encodes an inference problem for estimating the probability distribution over the bias of a coin, \emph{conditioned} on the outcome of four experimental coin flips: \true{}, \true{}, \false{}, and \true{} ($\true{} = \text{heads}$ and $\false{} = \text{tails}$).
At line~\ref{line:base:assume}, we use the PPL-specific \ttt{assume} construct to define our \emph{prior} belief in the bias $a_1$ of the coin.
We set this prior belief to a Beta$(2,2)$ probability distribution, illustrated in Fig.~\ref{fig:running:beta}.
In the illustration, 0 indicates a coin that always results in \false{}, 1 a coin that always results in \true{}, and 0.5 a fair coin.
We see that our prior belief is quite evenly spread out, but with more probability mass towards a fair coin.
To condition this prior distribution on the observed coin flips, we conceptually execute the program in Fig~\ref{fig:running:base} infinitely many times, \emph{sampling} values from the prior Beta distribution at \ttt{assume} (line~\ref{line:base:assume}) and, as a side effect, \emph{accumulating the product of weights} given as argument to the PPL-specific \ttt{weight} construct (line~\ref{line:base:weight}).
We make the four consecutive calls \ttt{weight ($f_\textrm{Bernoulli}$ $a_1$ $\true{}$)}, \ttt{weight ($f_\textrm{Bernoulli}$ $a_1$ $\true{}$)}, \ttt{weight ($f_\textrm{Bernoulli}$ $a_1$ $\false{}$)}, and \ttt{weight ($f_\textrm{Bernoulli}$ $a_1$ $\true{}$)}\footnote{%
  PPLs also commonly use a similar built-in function \ttt{observe} to update the weight. For example, \ttt{observe ($\textrm{Bernoulli}$ $a_1$) \true{}} is equivalent to \ttt{weight ($f_\textrm{Bernoulli}$ $a_1$ $\true{}$)}.
}, using the recursive function $\mi{iter}$.
The function application \ttt{$f_\textrm{Bernoulli}$ $a_1$ $o$} gives the probability of the outcome $o$ given a bias $a_1$ for the coin.
I.e., $\ttt{$f_\textrm{Bernoulli}$ $a_1$ $\true{}$} = a_1$ and $\ttt{$f_\textrm{Bernoulli}$ $a_1$ $\false{}$} = 1 - a_1$.
So, for example, a sample $a_1 = 0.4$ gets the accumulated weight $0.4\cdot0.4\cdot0.6\cdot0.4$ and $a_1 = 0.7$ the accumulated weight $0.7\cdot0.7\cdot0.3\cdot0.7$.
The end result is an infinite set of \emph{weighted} samples of $a_1$ (the program returns $a_1$ at line~\ref{line:base:return}) that approximate the \emph{posterior} or \emph{target} distribution of Fig.~\ref{fig:running:base}, illustrated in Fig~\ref{fig:running:posterior}.
Note that, because we observed three \true{} outcomes and only one \false{}, the weights shift the probability mass towards 1 and narrows it slightly as we are now more sure about the bias of the coin.
Increasing the number of experimental coin flips would make Fig.~\ref{fig:running:posterior} more and more narrow.

We can approximate the infinite number of samples by running the program a large (but finite) number of times.
This basic inference algorithm is known as \emph{likelihood weighting}.
The problem with likelihood weighting is that it is only accurate enough for simple models.
For complex models, it is common that only a few likelihood weighting samples (often only one) get much larger weights relative to the other samples, greatly reducing inference accuracy.
Real-world models require more powerful inference algorithms based on, e.g., SMC or MCMC.
A key requirement in both SMC and MCMC is the ability to \emph{suspend} executions of probabilistic programs at calls to \ttt{weight} and/or \ttt{assume}.
One way to enable suspensions is by writing programs in CPS.
We first illustrate a simple use of CPS to suspend at \ttt{assume} in Fig.~\ref{fig:running:assume}.
Here, the program immediately returns an object $\Susassume(\textrm{Beta}$~$2$~$2$,~$k)$, indicating that execution stopped at an \ttt{assume} with the argument \textrm{Beta}~$2$~$2$ and a continuation $k$ (i.e., the abstraction binding $a_1$) that executes the remainder of the program.
With likelihood weighting, we would simply sample a value $a_1$ from the \textrm{Beta}~$2$~$2$ distribution and resume execution by calling $k$~$a_1$.
This call then runs the program until termination and results in the actual return value of the program, which is $a_1$.
Many MCMC inference algorithms reuse samples from previous executions at $\Susassume$, and the suspensions are thus useful to avoid unnecessary re-execution~\cite{ritchie2016c3}.

As a second example, we illustrate suspension at \ttt{weight} for, e.g., SMC inference in Fig.~\ref{fig:running:weight}.
Here, we require suspensions in the middle of the recursive call to $\mi{iter}$, and writing the program in CPS is more challenging.
We rewrite the $\mi{iter}$ function to take a continuation $k$ as argument, and call the continuation with the return value $()$ at line~\ref{line:weight:ret} instead of directly returning $()$ as in Fig.~\ref{fig:running:base} at line~\ref{line:base:ret}.
This continuation argument $k$ is precisely what allows us to construct and return $\Susweight$ objects at line~\ref{line:weight:sus}.
To illustrate the suspensions, consider executing the program with likelihood weighting.
First, the program returns the object $\Susweight(f_{\textrm{Bernoulli}(a_1)}~\true{}, k')$, where $k'$ is the continuation that line~\ref{line:weight:k} constructs.
Likelihood weighting now updates the weight for the execution with the value $f_{\textrm{Bernoulli}(a_1)}~\true{}$ and resumes execution by calling $k'~()$.
Similarly, this next execution returns $\Susweight(f_{\textrm{Bernoulli}(a_1)}~\true{}, k'')$ for the second recursive call to $\mi{iter}$, and we again update the weight and resume by calling $k''~()$.
We similarly encounter $\Susweight(f_{\textrm{Bernoulli}(a_1)}~\false{}, k''')$ and $\Susweight(f_{\textrm{Bernoulli}(a_1)}~\true{}, k'''')$ before the final call $k''''~()$ runs the program to termination and produces the actual return value $a_1$.
In SMC, we run many executions concurrently and wait until they all have returned a $\Susweight$ object.
At this point, we resample the executions according to their weights (the first value in $\Susweight$), which discards executions with low weight and replicates executions with high weight.
After resampling, we continue to the next suspension and resampling by calling the continuations.

PPL implementations enable suspensions at \ttt{assume} and/or \ttt{weight} through automatic and full CPS transformations.
Fig.~\ref{fig:running:full} illustrates such a transformation for Fig.~\ref{fig:running:base}.
We indicate CPS versions of intrinsic functions with the $_\textrm{CPS}$ subscript.
Note that the full CPS transformation results in many additional closure allocations compared to Fig.~\ref{fig:running:assume} and Fig.~\ref{fig:running:weight}.
As a result, runtime overhead increases significantly.
The contribution in this paper is a static analysis that allows an automatic and selective CPS transformation of programs, as in Fig.~\ref{fig:running:assume} and Fig.~\ref{fig:running:weight}.
With a selective transformation, we avoid many unnecessary closure allocations, and can significantly reduce runtime overhead while still allowing suspensions as required for a given inference algorithm.

\section{Syntax and Semantics}\label{sec:ppl}
This section introduces the PPL calculus used to formalize the suspension analysis in Section~\ref{sec:sus} and selective CPS transformation in Section~\ref{sec:cps}.
Section~\ref{sec:syntax} gives the abstract syntax and Section~\ref{sec:semantics} a big-step operational semantics.
Section~\ref{sec:anf} introduces A-normal form---a prerequisite for both the suspension analysis and the selective CPS transformation.

\subsection{Syntax}\label{sec:syntax}
We build upon the standard untyped lambda calculus, representative of functional universal PPLs such as Anglican, WebPPL, and Miking CorePPL.
We define the abstract syntax below.
\begin{definition}[Terms, values, and environments]\label{def:terms}
  We define terms $\term \in T$ and values $\termv \in V$ as
  {\upshape
    \begin{equation}\label{eq:ast}
      \begin{gathered}
        \begin{aligned}
          \term \Coloneqq& \s
          x
          \s | \s
          c
          \s | \s
          \lambda x. \s \term
          \s | \s
          \term \s \term
          \s | \s
          \ttt{let } x = \term \ttt{ in } \term
          & \termv \Coloneqq& \s
          c
          \s | \s
          \langle\lambda x. \s \term,\rho\rangle
          \\ |& \s
          \ttt{if } \term \ttt{ then } \term \ttt{ else } \term
          \s | \s
          \ttt{assume } \term
          \s | \s
          \ttt{weight } \term
          &&\\
        \end{aligned} \\
        x,y \in X \quad
        \rho \in P \quad
        c \in C \quad
        \{ \false{}, \true{}, () \} \cup \mathbb{R} \cup D \subseteq C.
      \end{gathered}
    \end{equation}
  }%
  The countable set $X$ contains variable names, $C$ intrinsic values and operations, and $D \subset C$ intrinsic probability distributions.
  The set $P$ contains \emph{evaluation environments}, i.e., maps from variables in $X$ to values in $V$.
\end{definition}
\begin{definition}[Target language terms]
  As a target language for the selective CPS transformation in Section~\ref{sec:cps},
  we additionally extend Definition~\ref{def:terms} to target language terms $\term \in T^+$ by
  {\upshape
    \begin{equation}
      \term \mathrel{+}=
      \Susassume(\term,\term)
      \s | \s
      \Susweight(\term,\term).
    \end{equation}
  }
\end{definition}
\noindent
Fig.~\ref{fig:running:base} gives an example of a term in $T$, and Fig.~\ref{fig:running:assume} and Fig.~\ref{fig:running:weight} of terms in $T^+$.
However, note that the programs in Fig.~\ref{fig:running} also use the list constructor \ttt{[$\ldots$]} (not part of the above definitions) to make the example more interesting.

In addition to the standard variable, abstraction, and application terms in the untyped lambda calculus, we include explicit \ttt{let} expressions for convenience.
Furthermore, we use the syntactic sugar \ttt{let rec $f$ = $\lambda x.\term_1$ in $\term_2$} to define recursive functions (translating to an application of a call-by-value fixed-point combinator).
We use \ttt{$\term_1$; $\term_2$} as a shorthand for \ttt{($\lambda \_. \term_2$) $\term_1$}, where $\_$ means that we do not use the argument.
That is, we evaluate $\term_1$ for side effects only.

We include a set $C$ of intrinsic operations and constants essential to inference problems encoded in PPLs.
The set of intrinsics includes boolean truth values, the unit value, real numbers, and probability distributions.
We can also add further operations and constants to $C$.
For example, we can let $+ \in C$ to support addition of real numbers.
To allow control flow to depend on intrinsic values, we include \ttt{if} expressions that use intrinsic booleans as condition.

We saw examples of the \ttt{assume} and \ttt{weight} constructs in Section~\ref{sec:motivating}.
The \ttt{assume} construct takes distributions $D \subset C$ as argument, and produces random variables distributed according to these distributions.
For example, we can let $\mathcal N \in C$ be a function that constructs normal distributions.
Then, \ttt{assume ($\mathcal N$ $0$ $1$)}, where $\mathcal N$ $0$ $1 \in D$, defines a random variable with a standard normal distribution.
Partially constructed distributions, e.g., $\mathcal N$ $0$, are also in $C$, but not in $D$ (they are not yet proper distributions).
As we saw in Section~\ref{sec:motivating}, the \ttt{weight} construct updates the likelihood with the real number given as argument, and allows conditioning on data (e.g., the four coin flips in Fig.~\ref{fig:running}).

\subsection{Semantics}\label{sec:semantics}
We construct a call-by-value big-step operational semantics, based on Lundén et al.~\cite{lunden2023automatic}, describing how to evaluate terms $\term \in T$.
Such a semantics is a key component when formally defining the probability distributions corresponding to terms $\term \in T$ (e.g., the distribution in Fig.~\ref{fig:running:posterior} corresponding to the program in Fig.~\ref{fig:running:base}) and also when proving various properties of PPLs and their inference algorithms (e.g., inference correctness).
See, e.g., the work by Borgström et al.~\cite{borgstrom2016lambda} and Lundén et al.~\cite{lunden2021correctness} for full formal treatments.

\begin{figure}[tb]
  \[\footnotesize
    \begin{gathered}
      \frac{ \rho \vdash \term_1 \sem{u_1}{s_1}{w_1} \langle\lambda x. \term,\rho'\rangle \quad \rho \vdash \term_2 \sem{u_2}{s_2}{w_2} \termv_2 \quad \rho' ,x \mapsto \termv_2 \vdash \term \sem{u_3}{s_3}{w_3} \termv}
      { \rho \vdash \term_1 \s \term_2 \sem{u_1 \lor u_2 \lor u_3}{s_1 \concat s_2 \concat s_3}{w_1 \cdot w_2 \cdot w_3} \termv }
      (\textsc{App}) \\[0.5em]
      \frac{}
      { \rho \vdash x \sem{\false{}}{[]}{1} \rho(x) }
      (\textsc{Var}) \qquad
      \frac{ \rho \vdash \term_1 \sem{u_1}{s_1}{w_1} c_1 \quad \rho \vdash \term_2 \sem{u_2}{s_2}{w_2} c_2}
      { \rho \vdash \term_1 \s \term_2 \sem{u_1 \lor u_2}{s_1 \concat s_2}{w_1 \cdot w_2} \delta(c_1,c_2) }
      (\textsc{Const-App}) \\[0.5em]
      \frac{}
      { \rho \vdash \lambda x. \term \sem{\false{}}{[]}{1} \langle\lambda x. \term,\rho\rangle }
      (\textsc{Lam}) \qquad
      \frac{ \rho \vdash \term_1 \sem{u_1}{s_1}{w_1} \termv_1 \quad \rho, x \mapsto \termv_1 \vdash \term_2 \sem{u_2}{s_2}{w_2} \termv}
      { \rho \vdash \ttt{let } x = \term_1 \ttt{ in } \term_2 \sem{u_1 \lor u_2}{s_1 \concat s_2}{w_1 \cdot w_2} \termv }
      (\textsc{Let}) \\[0.5em]
      \frac{}
      { \rho \vdash c \sem{\false{}}{[]}{1} c }
      (\textsc{Const}) \qquad
      \frac{ \rho \vdash \term_1 \sem{u_1}{s_1}{w_1} \true{} \quad \rho \vdash \term_2 \sem{u_2}{s_2}{w_2} \termv_2 }
      {\rho \vdash \ttt{if } \term_1 \ttt{ then } \term_2 \ttt{ else } \term_3 \sem{u_1 \lor u_2}{s_1 \concat s_2}{w_1 \cdot w_2} \termv_2}
      (\textsc{If-True}) \\[0.5em]
      \frac{\rho \vdash \term \sem{u}{s}{w} d \quad w' = f_d(c) }
      {\rho \vdash \ttt{assume } \term \sem{\suspend_\ttt{assume} \lor u}{s \concat [c]}{w \cdot w'} c}
      (\textsc{Assume}) \hspace{0.5mm}
      \frac{\rho \vdash \term \sem{u}{s}{w} w'}
      {\rho \vdash \ttt{weight } \term \sem{\suspend_\ttt{weight} \lor u}{s}{w \cdot w'} ()}
      (\textsc{Weight}) \\[0.5em]
    \end{gathered}
  \]
  \caption{%
    A big-step operational semantics for $\term \in T$.
    We omit the rule (\textsc{If-False}) for brevity; it is analogous to (\textsc{If-True}).
    The environment $\rho, x \mapsto \termv$ denotes $\rho$ extended with a binding $\termv$ for $x$.
    For each $d \in D$, the function $f_d$ is its probability density or probability mass function.
    E.g., $f_{\mathcal N(0,1)}(x) = e^{x^2/2}/\sqrt{2\pi}$, the density function of the standard normal distribution.
    We use the following notation: $\concat$ for sequence concatenation, $\cdot$ for multiplication, and $\lor$ for logical disjunction.
  }
  \label{fig:semantics}
\end{figure}

We use the semantics to formally define suspension, and use this definition to state the soundness of the suspension analysis in Section~\ref{sec:sus} (Theorem~\ref{thm:main}).
We use a big-step semantics, as we do not require the additional control provided by a small-step semantics.
For example, we do not concern ourselves with details of termination, as the soundness of the analysis relates only to terminating executions.
Fig.~\ref{fig:semantics} presents the full semantics as a relation $\rho \vdash \term \sem{u}{s}{w} \termv$ over tuples $(P, T, S, \{\false{},\true{}\}, \mathbb{R}, V)$.
$S$ is a set of \emph{traces} capturing the random draws at \ttt{assume} during evaluation.
Intuitively, $\rho \vdash \term \sem{u}{s}{w} \termv$ holds iff $\term$ evaluates to $\termv$ in the environment $\rho$ with the trace $s$ and the total probability density (i.e., the accumulated weight) $w$.
We describe the suspension flag $u$ later in this section.

Most of the rules are standard and we focus on explaining key properties related to PPLs and suspension.
We first consider the rule (\tsc{Const-App}), which uses the $\delta$-function to evaluate intrinsic operations.
\noindent
\begin{definition}[Intrinsic arities and the $\delta$-function]\label{def:const}
  For each $c \in C$, we let $|c| \in \mathbb N$ denote its \emph{arity}.
  We also assume the existence of a partial function $\delta: C \times C \rightarrow C$ such that if $\delta(c, c_1) = c_2$, then $|c| > 0$ and $|c_2| = |c| - 1$.
\end{definition}
\noindent
For example, $\delta((\delta(+,1)), 2) = 3$.
We use the arity property of intrinsics to formally define traces.
\begin{definition}[Traces]\label{def:trace}
  For all $s \in S$, $s$ is a sequence of intrinsics with arity 0, called a \emph{trace}.
  We write $s = [c_1,c_2,\ldots,c_n]$ to denote a trace $s$ with $n$ elements.
\end{definition}
\noindent
The rule (\tsc{Assume}) formalizes random draws and consumes elements of the trace.
Specifically, (\tsc{Assume}) updates the evaluation's total probability density $w \in \mathbb R$ with the density $w'$ of the first trace element with respect to the distribution given as argument to \ttt{assume}.
The rule (\tsc{Weight}) furthermore directly modifies the total probability density according to the \ttt{weight} argument.

We now consider the special suspension flag $u$ in the derivation $\rho \vdash \term \sem{u}{s}{w} \termv$.
\begin{definition}[Suspension requirement]\label{def:sus}
  A derivation $\rho \vdash \term \sem{u}{s}{w} \termv$ \emph{requires suspension} if the suspension flag $u$ is $\true{}$.
\end{definition}
For example, the rule (\tsc{App}) requires suspension if $u_1 \lor u_2 \lor u_3$---i.e., if any subderivation requires suspension.
To reflect the particular suspension requirements in SMC and MCMC inference, we limit the source of suspension requirements to \ttt{assume} and \ttt{weight}.
We turn the individual sources on and off through the boolean variables $\suspend_\ttt{assume}$ and $\suspend_\ttt{weight}$ in Fig.~\ref{fig:semantics}.
For the examples in the remainder of this paper, we let $\suspend_\ttt{weight} = \true{}$ and $\suspend_\ttt{assume} = \false{}$ (i.e., only \ttt{weight} requires suspension, as in SMC inference).

To illustrate the semantics, consider $\texample$ of Fig.~\ref{fig:running:base} again.
Because $\texample$ evaluates precisely one \ttt{assume}, the only valid traces for $\texample$ are singleton traces $[a_1]$, where $a_1 \in \mathbb R_{[0,1]}$ due to the Beta prior for $a_1$.
By initially setting $\rho$ to the empty environment $\varnothing$ and following the rules of Fig.~\ref{fig:semantics}, we derive
$
  \varnothing \vdash \texample \sem{\true{}}{[a_1]}{f_{\textrm{Beta}(2,2)}(a_1)\cdot a_1^3(1-a_1)} a_1.
$
Note that every evaluation of $\texample$ has $u = \true{}$, as there are always four calls to \ttt{weight} during evaluation.
That is, the derivation requires suspension.
However, many subderivations of $\texample$ do \emph{not} require suspension.
For example, the subderivations \ttt{assume (\textrm{Beta} $2$ $2$)} and \ttt{$\mi{null}$ $\mi{obs}$} do not (i.e., have $u = \false$).
Section~\ref{sec:sus} presents a suspension analysis that conservatively approximates which subderivations require suspension.
The analysis enables, e.g., the selective CPS transformation in Fig.~\ref{fig:running:weight}.

\subsection{A-Normal Form}\label{sec:anf}
We simplify the suspension analysis in Section~\ref{sec:sus} and the selective CPS transformation in Section~\ref{sec:cps} by requiring that terms are in \emph{A-normal form} (ANF)~\cite{flanagan1993essence}.
\begin{definition}[A-normal form]
  We define the A-normal form terms $\termanf \in \termanfty$ as follows.
  {\upshape
    \begin{equation}\label{eq:anf}
      \begin{aligned}
        &\begin{aligned}
          \termanf \Coloneqq& \s
          x
          \s | \s
          \ttt{let } x = \termanf' \ttt{ in } \termanf
          \\
          \termanf' \Coloneqq& \s
          x
          \s | \s
          c
          \s | \s
          \lambda x. \s \termanf
          \s | \s
          x \s y
          \\ |& \s
          \ttt{if } x \ttt{ then } \termanf \ttt{ else } \termanf
          \s | \s
          \ttt{assume } x
          \s | \s
          \ttt{weight } x
        \end{aligned}
      \end{aligned}
    \end{equation}
  }
\end{definition}
\begin{figure}[tb]
  \centering
  \lstset{%
    basicstyle=\ttfamily\scriptsize,
    showlines=true,
    framexleftmargin=-2pt,
    xleftmargin=2em,
  }%
  \begin{multicols}{2}
    \begin{lstlisting}[style=ppl]
let $t_1$ = $2$ in
let $t_2$ = $2$ in
let $t_3$ = $\textrm{Beta}$ in
let $t_4$ = $t_3$ $t_1$ in
let $t_5$ = $t_4$ $t_2$ in
let $a_1$ = assume $t_5$ in
let rec $\mi{iter}$ = $\lambda \mi{obs}.$$\label{line:anflam}$
  let $t_6$ = $\mi{null}$ in
  let $t_7$ = $t_6$ $\mi{obs}$ in
  let $t_8$ = $\label{line:if}$
    if $t_7$ then
      let $t_9$ = $()$ in
      $t_9$
    else
      let $t_{10}$ = $f_\textrm{Bernoulli}$ in
      let $t_{11}$ = $t_{10}$ $a_1$ in
      let $t_{12}$ = $\mi{head}$ in
      let $t_{13}$ = $t_{12}$ $\mi{obs}$ in
      let $t_{14}$ = $t_{11}$ $t_{13}$ in
      let $w_1$ = weight $t_{14}$ in$\label{line:weight}$
      let $t_{15}$ = $\mi{tail}$ in
      let $t_{16}$ = $t_{15}$ $\mi{obs}$ in
      let $t_{17}$ = $\mi{iter}$ $t_{16}$ in$\label{line:iter1}$
      $t_{17}$
  in
  $t_8$$\label{line:obsreturn}$
in
let $t_{18}$ = $\true{}$ in
let $t_{19}$ = $\false{}$ in
let $t_{20}$ = $\true{}$ in
let $t_{21}$ = $\true{}$ in
let $t_{22}$ = $[t_{21}$,$t_{20}$,$t_{19}$,$t_{18}]$ in
let $t_{23}$ = $\mi{iter}$ $t_{22}$ in$\label{line:iter2}$
$a_1$$\label{line:a1}$
    \end{lstlisting}
  \end{multicols}

  \caption{
    The running example $\texample$ from Fig.~\ref{fig:running:base} transformed to ANF.
  }
  \label{fig:runninganf}
\end{figure}
\noindent
It holds that $\termanfty \subset T$.
Furthermore, there exist standard transformations to convert terms in $T$ to $\termanfty$.
Fig.~\ref{fig:runninganf} illustrates Fig.~\ref{fig:running:base} transformed to ANF.
We will use Fig.~\ref{fig:running:base} as a running example in Section~\ref{sec:sus} and Section~\ref{sec:cps}.

Restricting programs to ANF significantly simplifies the suspension analysis and selective CPS transformation.
From now on we require that all variable bindings in programs are unique, and together with ANF, the result is that every expression in a program $\term \in \termanfty$ is \emph{uniquely labeled} by a variable name from a \ttt{let} expression.
This property is essential for the treatment in Section~\ref{sec:sus}.

\section{Suspension Analysis}\label{sec:sus}
This section presents the main technical contribution: the suspension analysis.
The analysis goal is to identify program expressions that may require suspension in the sense of Definition~\ref{def:sus}.
Identifying such expressions leads to the selective CPS transformation in Section~\ref{sec:cps}, enabling transformations such as in Fig~\ref{fig:running:weight}.

The suspension analysis builds upon the 0-CFA algorithm~\cite{shivers1991control,nielson1999principles}, and we formalize our algorithms based on Lundén et al.~\cite{lunden2023automatic}.
The main challenge we solve is how to model the propagation of suspension in the presence of higher-order functions.
The 0 in 0-CFA stands for \emph{context insensitivity}---the analysis considers every part of the program in one global context.
Context insensitivity makes the analysis more conservative compared to context-sensitive approaches such as $k$-CFA, where $k \in \mathbb N$ indicates the level of context sensitivity~\cite{midtgaard2021control}.
We use 0-CFA for two reasons: (i) the worst-case time complexity for the analysis is polynomial, while it is exponential for $k$-CFA already at $k = 1$, and (ii) the limitations of 0-CFA rarely matter in practical PPL applications.
For example, $k$-CFA provides no benefits over 0-CFA for the programs in Section~\ref{sec:evaluation}.

We assume
$\langle\lambda x. \s \term, \rho \rangle \not\in C$ (recall that $C$ is the set of intrinsics).
That is, we assume that closures are not part of the intrinsics.
In particular, this disallows intrinsic operations (including the use of \ttt{assume $d$}, $d \in D \subset C$) to produce closures, which would needlessly complicate the analysis without any benefit.

Consider the program in Fig.~\ref{fig:runninganf}, and assume that \ttt{weight} requires suspension.
Clearly, the expression labeled by $w_1$ at line~\ref{line:weight} then requires suspension.
Furthermore, $w_1$ evaluates as part of the larger expression labeled by $t_8$ at line~\ref{line:if}.
Consequently, the evaluation of $t_8$ also requires suspension.
Also, $t_8$ evaluates as part of an application of the abstraction binding $\mi{obs}$ at line~\ref{line:anflam}.
In particular, the abstraction binding $\mi{obs}$ binds to $\mi{iter}$, and we apply $\mi{iter}$ at lines~\ref{line:iter1} and~\ref{line:iter2}.
Thus, the expressions named by $t_{17}$ and $t_{22}$ require suspension.
In summary, we have that $w_1$, $t_8$, $t_{17}$, and $t_{22}$ require suspension, and we also note that all applications of the abstraction binding $\mi{obs}$ require suspension.

We proceed to the formalization and first introduce standard \emph{abstract values}.
\begin{definition}[Abstract values]\label{def:absval}
  We define the abstract values {\upshape$\absval \in A$} as
  {\upshape$
    \absval \Coloneqq
    \lambda x. y
    \s | \s
    \ttt{const}_x \, n
  $} \hspace{2mm}for $x,y \in X$ and $n \in \mathbb{N}$.
\end{definition}
\noindent
The abstract value $\lambda x. y$ represents all closures originating at, e.g., a term \ttt{$\lambda x.$ let $y$ = $1$ in $y$} in a program at runtime (recall that we assume that the variables $x$ and $y$ are unique).
Note that the $y$ indicates the name returned by the body (formalized by the function \tsc{name} in Algorithm~\ref{alg:gencstr}).
The abstract value $\ttt{const}_x \, n$ represents all intrinsic functions of arity $n$ originating at $x$.
For example, $\ttt{const}_x \s 2$ originates at, e.g., a term \ttt{let $x$ = $+$ in \term}.

The central objects in the analysis are sets $S_x \in \mathcal P (A)$ and boolean values $\suspend_x$ for all $x \in X$.
The set $S_x$ contains all abstract values that may flow to the expression labeled by $x$, and $\suspend_x$ indicates whether or not the expression requires suspension.
A trivial but useless solution is $S_x = A$ and $\suspend_x = \true{}$ for all variables $x$ in the program.
To get more precise information regarding suspension, we wish to find smaller solutions to the $S_x$ and $\suspend_x$.

\begin{algorithm}[tb]
  \renewcommand{\s}{\hphantom{|}}
  \caption{%
    Constraint generation for the suspension analysis.
    We write the functional-style pseudocode for the algorithm itself in sans serif font to distinguish it from terms in $T$.
  }\label{alg:gencstr}
  \raggedright
  \lstinline[style=alg]!function $\tsc{generateConstraints}$($\term$): $\termanfty \rightarrow \mathcal P(R)$ =!\\
  \hspace{3mm}
  \begin{minipage}{0.96\textwidth}
    \begin{multicols}{2}
      \begin{lstlisting}[
          style=alg,
          showlines=true,
          basicstyle=\sffamily\scriptsize,
      ]
match $\term$ with$\label{line:topmatch}$
| $x \rightarrow$ $\varnothing$
| $\ttt{let } x = \term_1 \s \ttt{in} \s \term_2 \rightarrow$$\label{line:beginlet}$
$\s$ $\tsc{generateConstraints}(\term_2) \s \cup$
  $\s$ match $\term_1$ with
  $\s$ | $y \rightarrow \{S_y \subseteq S_x\}$$\label{line:genvar}$
  $\s$ | $c \rightarrow$ if $|c| > 0$ then $\{\ttt{const}_x \s |c|\in S_x\}$
  $\s$ $\s$ $\hspace{5mm}$ else $\varnothing$
  $\s$ | $\lambda y. \s \term_b \rightarrow$ $\tsc{generateConstraints}(\term_b)$$\label{line:genabsb}$
  $\s$ $\s$ $\s$ $\cup \s \{\lambda y. \s \tsc{name} \s \term_b \in S_x \}$
  $\s$ $\s$ $\s$ $\cup \s \{ \suspend_n \Rightarrow \suspend_y$
  $\hspace{12mm} \mid n \in \tsc{suspendNames}(t_b)\}$$\label{line:genabse}$
  $\s$ | $\mi{lhs} \s \mi{rhs} \rightarrow$ $\{$$\label{line:genappb}$
    $\s$ $\s$ $\forall z \forall y \s \lambda z. y \in S_{\mi{lhs}}$
    $\s$ $\s$ $\s$ $\Rightarrow (S_\mi{rhs} \subseteq S_z) \land (S_y \subseteq S_x)$,
    $\s$ $\s$ $\forall y \forall n \s \ttt{const}_y \, n \in S_\mi{lhs} \land n > 1 $
    $\s$ $\s$ $\s$ $\Rightarrow \ttt{const}_y \, n-1 \in S_x$,
    $\s$ $\s$ $\forall y \s \lambda y. \_ \in S_{\mi{lhs}}$
    $\s$ $\s$ $\s$ $\Rightarrow (\suspend_y \Rightarrow \suspend_x)$,
    $\s$ $\s$ $\forall y \s \ttt{const}_y \s \_ \in S_{\mi{lhs}}$
    $\s$ $\s$ $\s$ $\Rightarrow (\suspend_y \Rightarrow \suspend_x)$,
    $\s$ $\s$ $\suspend_x \Rightarrow$
    $\s$ $\s$ $\s$ $(\forall y \s \lambda y. \_ \in S_{\mi{lhs}} \Rightarrow \suspend_y)$
    $\s$ $\s$ $\s$ $\land \s (\forall y \s \ttt{const}_y \s \_ \in S_{\mi{lhs}} \Rightarrow \suspend_y)$
  $\s$ $\s$ $\}$$\label{line:genappe}$
  $\s$ | $\ttt{assume } \textrm{\_} \rightarrow$
  $\s$ $\s$ if $\suspend_\ttt{assume}$ then $\{ \suspend_x \}$ else $\varnothing$
  (*@ \columnbreak @*)
  $\s$ | $\ttt{weight } \textrm{\_} \rightarrow$
  $\s$ $\s$ if $\suspend_\ttt{weight}$ then $\{ \suspend_x \}$ else $\varnothing$ $\label{line:genweight}$
  $\s$ | $\ttt{if } y \ttt{ then } \term_t \ttt{ else } \term_e \rightarrow$$\label{line:genifb}$
  $\s$ $\s$ $\tsc{generateConstraints}(\term_t)$
    $\s$ $\s$ $\cup \s \tsc{generateConstraints}(\term_e)$
    $\s$ $\s$ $\cup \s \{S_{\tsc{name} \s \term_t} \subseteq S_x, S_{\tsc{name} \s \term_e} \subseteq S_x\}$
    $\s$ $\s$ $\cup \s \{\suspend_n \Rightarrow \suspend_x $
           $\mid n \in \tsc{suspendNames}(\term_t)$
                  $\cup \s \tsc{suspendNames}(\term_e)\}$$\label{line:genife}$

function $\tsc{name}$($\term$): $\termanfty \rightarrow X$ =$\label{line:func_name}$
  match $\term$ with
  | $x \rightarrow x$
  | $\ttt{let } x = \term_1 \s \ttt{in} \s \term_2 \rightarrow$ $\tsc{name}$($\term_2$)

function $\tsc{suspendNames}$($\term$): $\termanfty \rightarrow \mathcal P(X)$ = $\label{line:func_suspendNames}$
  match $\term$ with
  | $x \rightarrow \varnothing$
  | $\ttt{let } x = \term_1 \s \ttt{in} \s \term_2 \rightarrow$
  $\s$ $\tsc{suspendNames}(\term_2) \s \cup$
    $\s$ match $\term_1$ with
    $\s$ | $\mi{lhs} \s \mi{rhs} \rightarrow \{ x \}$
    $\s$ | $\ttt{if } y \ttt{ then } \term_t \ttt{ else } \term_e \rightarrow \{ x \}$
    $\s$ | $\ttt{assume}$ $\_$ $\rightarrow$
    $\s$ $\s$ if $\suspend_\ttt{assume}$ then $\{ x \}$ else $\varnothing$
    $\s$ | $\ttt{weight}$ $\_$ $\rightarrow$
    $\s$ $\s$ if $\suspend_\ttt{weight}$ then $\{ x \}$ else $\varnothing$
    $\s$ | $\_ \rightarrow$ $\varnothing$
      \end{lstlisting}
    \end{multicols}
  \end{minipage}
\end{algorithm}%
To formalize the set of sound solutions for $S_x$ and $\suspend_x$, we generate \emph{constraints} $\cstr \in R$ for programs\ifextended\ (for a formal definition of constraints, see Appendix~\ref{sec:cfaalg}). \fi
Algorithm~\ref{alg:gencstr} formalizes the necessary constraints for programs $\term \in \termanfty$ with a function \tsc{generateConstraints} that recursively traverses the program $\term$ to generate a set of constraints.
Due to ANF, there are only two cases in the top match (line~\ref{line:topmatch}).
Variables generate no constraints, and the important case is for \ttt{let} expressions at lines~\ref{line:beginlet}--\ref{line:genweight}.
The algorithm makes use of an auxiliary function \tsc{name} (line~\ref{line:func_name}) that determines the name of an ANF expression, and a function \tsc{suspendNames} (line~\ref{line:func_suspendNames}) that determines the names of all top-level expressions within an expression that may suspend (namely, applications, \ttt{if} expressions, and \ttt{assume} and/or \ttt{weight}).

We next illustrate and motivate the generated constraints by considering the set of constraints $\tsc{generateConstraints}(\texample)$, where $\texample$ is the program in Fig.~\ref{fig:runninganf}.
Many constraints are standard, and we therefore focus on the new suspension constraints introduced as part of this paper.
In particular, the challenge is to correctly capture the flow of suspension requirements across function applications and higher-order functions.
First, we see that defining aliases (line~\ref{line:genvar}) generates constraints of the form $S_y \subseteq S_x$, that constants introduce \ttt{const} abstract values (e.g., $\ttt{const}_{t_6} 1 \in S_{t_6}$), and that \ttt{assume} and \ttt{weight} introduce suspension requirements, e.g., $\suspend_{w_1}$ (shorthand for $\suspend_{w_1} = \true{}$).

First, we consider the constraints generated for $\lambda \mi{obs}.$ (line~\ref{line:anflam} in Fig.~\ref{fig:runninganf}) through the case at lines~\ref{line:genabsb}-\ref{line:genabse} in Algorithm~\ref{alg:gencstr}.
To keep the example simple, we treat the unexpanded \ttt{let rec} as an ordinary \ttt{let} in the analysis (for this particular example, the analysis result is unaffected).
Omitting the recursively generated constraints for the abstraction body, the generated constraints are
\begin{equation}
  \{ \lambda \mi{obs}. \, t_8 \in S_\mi{iter}\} \cup \{ \suspend_n \Rightarrow \suspend_\mi{obs} \mid n \in \{ t_7, t_8\}
  \}.
\end{equation}
The first constraint is standard and states that the abstract value $\lambda \mi{obs}. \, t_8$ flows to $S_\mi{iter}$ as the variable naming the $\lambda \mi{obs}$ expression is $t_8$ at line~\ref{line:obsreturn} in Fig.~\ref{fig:runninganf} (difficult to notice due to the column breaks).
The remaining constraints are new and sets up the flow of suspension requirements.
Specifically, the abstraction $\mi{obs}$ itself requires suspension if any expression bound by a top-level \ttt{let} in its body requires suspension.
For efficiency, we only set up dependencies for expressions that may suspend (formalized by \tsc{suspendNames} in Algorithm~\ref{alg:gencstr}).
Note here that we do not add the constraint $\suspend_{w_1} \Rightarrow \suspend_\mi{obs}$, as $w_1$ is not at top-level in the body of $\mi{obs}$.
Instead, we later add the constraint $\suspend_{w_1} \Rightarrow \suspend_{t_8}$, and $\suspend_{w_1} \Rightarrow \suspend_\mi{obs}$ follows by transitivity.

The constraints generated for the \ttt{if} bound to $t_8$ at line~\ref{line:if} through the case at lines~\ref{line:genifb}-\ref{line:genife} in Algorithm~\ref{alg:gencstr} are (omitting recursively generated constraints)
\begin{equation}
  \begin{aligned}
    &\{ S_{t_9} \subseteq S_{t_8}, S_{t_{17}} \subseteq S_{t_8}\} \\
    &\hspace{10mm}\cup \{ \suspend_n \Rightarrow \suspend_{t_8} \mid n \in \{ t_{11}, t_{13}, t_{14}, w_1, t_{16}, t_{17}\} \}.
  \end{aligned}
\end{equation}
The first two constraints are standard, and state that abstract values in the results of both branches flow to the result $S_{t_8}$.
The last set of constraints is new and similar to the abstraction suspension constraints.
The constraints capture that all expressions at top-level in both branches that require suspension also cause $t_8$ to require suspension.

Consider the application at line~\ref{line:iter1} in Fig.~\ref{fig:runninganf}.
The generated constraints through the case at lines~\ref{line:genappb}-\ref{line:genappe} in Algorithm~\ref{alg:gencstr} are
\begin{equation}\label{eq:appcstrs}
  \begin{aligned}
    \{ \s &\forall z \forall y \s \lambda z. y \in S_{\mi{iter}} \Rightarrow (S_{t_{16}} \subseteq S_z) \land (S_y \subseteq S_{t_{17}}),\\
      &\forall y \forall n \s \ttt{const}_y \, n \in S_\mi{iter} \land n > 1 \Rightarrow \ttt{const}_y \, n-1 \in S_{t_{17}},\\
      &\forall y \s \lambda y. \_ \in S_{\mi{iter}} \Rightarrow (\suspend_y \Rightarrow \suspend_{t_{17}}),\\
      &\forall y \s \ttt{const}_y \s \_ \in S_{\mi{iter}} \Rightarrow (\suspend_y \Rightarrow \suspend_{t_{17}}),\\
    &\suspend_{t_{17}} \Rightarrow (\forall y \s \lambda y. \_ \in S_{\mi{iter}} \Rightarrow \suspend_y) \\
  & \hspace{25mm}\land (\forall y \s \ttt{const}_y \s \_ \in S_{\mi{iter}} \Rightarrow \suspend_y) \s \}.
  \end{aligned}
\end{equation}
The first two constraints are standard and state how abstract values flow as a result of applications.
The last three constraints are new and relate to suspension.
The third and fourth constraints state that if an abstraction or intrinsic requiring suspension flows to $\mi{iter}$, the result $t_{17}$ of the application also requires suspension.
The fifth constraint states that if the result $t_{17}$ requires suspension, then \emph{all} abstractions and constants flowing to $\mi{iter}$ require suspension.
This last constraint is not strictly required to later prove the soundness of the analysis in Theorem~\ref{thm:main}, but, as we will see in Section~\ref{sec:cps}, it is required for the selective CPS transformation.

\ifextended
We find a solution to the constraints through Algorithm~\ref{alg:flow} in Appendix~\ref{sec:cfaalg}.
The algorithm propagates abstract values according to the constraints until fixpoint, and is fairly standard.
\fi
However, we extend the algorithm to support the new suspension constraints.
The algorithm is a function \tsc{analyzeSuspend}: $\termanfty \rightarrow ((X \rightarrow \mathcal P (A)) \times \mathcal P(X))$.
The function returns a map $\tsf{data}: X \rightarrow \mathcal P (A)$ that assigns sets of abstract values to all $S_x$ and a set $\tsf{suspend}: \mathcal P(X)$ that assigns $\suspend_x = \true$ iff $x \in \tsf{suspend}$.
Importantly, the assignments to $S_x$ and $\tsf{suspend}_x$ satisfy all generated constraints.
To illustrate the algorithm, here are the analysis results $\tsc{analyzeSuspend}(\texample)$:
\begin{equation}\label{eq:analyzerunning}
  \begin{gathered}
    \begin{gathered}
      S_\mi{iter} = \{ \lambda \mi{obs}. t_8 \} \quad
      S_{t_6} = \{ \ttt{const}_{t_6} 1 \} \quad
      S_{t_{10}} = \{ \ttt{const}_{t_{10}} 2 \} \\
      S_{t_{11}} = \{ \ttt{const}_{t_{10}} 1 \} \quad
      S_{t_{12}} = \{ \ttt{const}_{t_{12}} 1 \} \quad
      S_{t_{15}} = \{ \ttt{const}_{t_{15}} 1 \} \\
      S_n = \varnothing \mid \text{all other $n \in X$}
    \end{gathered} \\
    \begin{aligned}
      \suspend_n &= \true{} \mid n \in \{ \mi{obs}, w_1, t_8, t_{17}, t_{22} \} \\
      \suspend_n &= \false{} \mid \text{all other $n \in X$}.
    \end{aligned}
  \end{gathered}
\end{equation}
The above results confirm our earlier reasoning: the expressions labeled by $\mi{obs}$, $w_1$, $t_8$, $t_{17}$, and $t_{22}$ may require suspension.

We now consider the soundness of the analysis.
First, the soundness of 0-CFA is well established (see, e.g., Nielson et al.~\cite{nielson1999principles}) and extends to our new constraints, and we take the following lemma to hold without proof.
\begin{lemma}[0-CFA soundness]\label{lemma:cfa}
  For every $\term \in \termanfty$, the solution given by
  $
    \tsc{analyzeSuspend}(\term)
  $
  for $S_x$ and $\suspend_x$, $x \in X$, satisfies the constraints $ \tsc{generateConstraints}(\term)$.
\end{lemma}
\noindent
Next, we must show that the constraints themselves are sound.
Consider the evaluation of an arbitrary term $\term \in \termanfty$.
For each subderivation of $\term$, labeled by a name $x$ (due to ANF), it must hold that $\suspend_x = \true$ if the subderivation requires suspension.
Otherwise, the analysis is unsound.
Theorem~\ref{thm:main} formally captures the soundness.
Note that the analysis is conservative (i.e., incomplete), because it may find $\suspend_x = true$ even if the subderivation for $x$ does not require suspension.
\begin{theorem}[Suspension analysis soundness]\label{thm:main}
  Let $\term \in \termanfty$, $s \in S$, $u \in \{\false{},\true{}\}$,
  $w \in \mathbb{R}$, and $\termv \in V$ such that
  $
    \varnothing \vdash \term \sem{u}{s}{w} \termv.
  $
  Now, let $S_x$ and $\suspend_x$ for $x \in X$ according to $\tsc{analyzeSuspend}(\term)$.
  For every subderivation $(\rho \vdash \ttt{let } x = \term_1 \ttt{ in } \term_2 \sem{u_1 \lor u_2}{s_1 \concat s_2}{w_1 \cdot w_2} \termv')$ of $(\varnothing \vdash \term \sem{u}{s}{w} \termv)$, $u_1 = \true{}$ implies $\suspend_x = \true{}$.
\end{theorem}
\ifextended
\begin{proof}
  Follows directly by Lemma~\ref{lemma:suslemma} in Appendix~\ref{sec:proof} with $\rho = \varnothing$.
\end{proof}
\else
\fi
\noindent
The proof \ifextended of Lemma~\ref{lemma:suslemma} \fi uses Lemma~\ref{lemma:cfa} and structural induction over the derivation $\varnothing \vdash \term \sem{u}{s}{w} \termv$.

Next, we use the suspension analysis to selectively CPS transform programs.

\section{Selective CPS Transformation}\label{sec:cps}
\begin{algorithm}[tb]
  \renewcommand{\s}{\hphantom{|}}
  \caption{%
    Selective continuation-passing style transformation.
    We define $\termid = \lambda x. x$.
    The term $c_\textrm{CPS}$ is the CPS version of $c$.
    We write the functional-style pseudocode for the algorithm itself in sans serif font to distinguish it from terms in $T$.
  }\label{alg:cps}
  \raggedright
  \lstinline[style=alg]!function $\tsc{cps}$(vars, $\term$): $\mathcal P(X) \times \termanfty \rightarrow T^+$ =!\\
  \hspace{3mm}
  \begin{minipage}{0.96\textwidth}
    \begin{multicols}{2}
      \begin{lstlisting}[
          style=alg,
          showlines=true,
          basicstyle=\sffamily\scriptsize,
      ]
return $\tsc{cps}'$($\termid$, $\term$)$\label{line:cps:init}$

function $\tsc{cps}'$(cont,$\term$): $T \times \termanfty \rightarrow T^+$ =
  match $\term$ with$\label{line:cps:topmatch}$
  | $x \rightarrow$ if cont $= \termid$ then $\term$ else cont $\term$$\label{line:cps:casevar}$
  | $\ttt{let } x = \term_1 \s \ttt{in} \s \term_2 \rightarrow$$\label{line:cps:beginlet}$
  $\s$ let $\term_2' = \tsc{cps}'(\tsf{cont},\term_2)$ in
  $\s$ match $\term_1$ with
  $\s$ | $y \rightarrow$ $\ttt{let } x = \term_1 \s \ttt{in} \s \term_2'$
  $\s$ | $c \rightarrow$ $\ttt{let } x =$
           (if $x \in \tsf{vars}$ then $c_\textrm{CPS}$ else $c$) $\ttt{in} \s \term_2'$
  $\s$ | $\lambda y. \s \term_b \rightarrow$$\label{line:cps:abs}$
  $\s$ $\s$ let $\term_1'$ = if $y \in \tsf{vars}$
  $\s$ $\s$ $\s$ then $\lambda k. \lambda y. \s \tsc{cps}'(k,\term_b)$
  $\s$ $\s$ $\s$ else $\lambda y. \s \tsc{cps}'(\termid,\term_b)$
  $\s$ $\s$ in
  $\s$ $\s$ $\ttt{let } x = \term_1' \s \ttt{in} \s \term_2'$
  $\s$ | $\mi{lhs} \s \mi{rhs} \rightarrow$$\label{line:cps:app}$
  $\s$ $\s$  if $x \in \tsf{vars}$ then
  $\s$ $\s$ $\s$  if $\tsc{tailCall}(\term)$
  $\s$ $\s$ $\s$ then $\mi{lhs}$ cont $\mi{rhs}$
  $\s$ $\s$ $\s$ else $\mi{lhs}$ $(\lambda x. \term_2')$ $\mi{rhs}$
  $\s$ $\s$  else $\ttt{let } x = \term_1 \s \ttt{in} \s \term_2'$



  (*@ \columnbreak @*)
  $\s$ | $\ttt{if } y \ttt{ then } \term_t \ttt{ else } \term_e \rightarrow$$\label{line:cps:if}$
  $\s$ $\s$  if $x \in \tsf{vars}$ then
  $\s$ $\s$ $\s$  if $\tsc{tailCall}(\term)$ then
  $\s$ $\s$ $\s$ $\s$ $\ttt{if } y \ttt{ then } \tsc{cps}'(\tsf{cont},\term_t)$
  $\s$ $\s$ $\s$ $\s$ $\ttt{else } \tsc{cps}'(\tsf{cont},\term_e)$
  $\s$ $\s$ $\s$ else
  $\s$ $\s$ $\s$ $\s$ $\ttt{let } k = \lambda x. \term_2' \s \ttt{in} \s$
  $\s$ $\s$ $\s$ $\s$ $\ttt{if } y \ttt{ then } \tsc{cps}'(k,\term_t) \ttt{ else } \tsc{cps}'(k,\term_e)$
  $\s$ $\s$ else $\ttt{let } x = \ttt{if } y \ttt{ then } \tsc{cps}'(\termid,\term_t)$
  $\s$ $\s$ $\hspace{15.5mm}$ $\ttt{else } \tsc{cps}'(\termid,\term_e) \s \ttt{in} \s \term_2'$
  $\s$ | $\ttt{assume } y \rightarrow$ $\ttt{let } x = \term_1 \s \ttt{in} \s \term_2'$
  $\s$ $\s$ if $x \in \tsf{vars}$ then
  $\s$ $\s$ $\s$ if $\tsc{tailCall}(\term)$
  $\s$ $\s$ $\s$ then $\Susassume$($y$, cont)
  $\s$ $\s$ $\s$ else $\Susassume$($y$,$\lambda x. \tsc{cps}'(\tsf{cont},\term_2)$)
  $\s$ $\s$ else $\ttt{let } x = \term_1 \s \ttt{in} \s \term_2'$
  $\s$ | $\ttt{weight } y \rightarrow$ $\ttt{let } x = \term_1 \s \ttt{in} \s \term_2'$$\label{line:cps:weight}$
  $\s$ $\s$ if $x \in \tsf{vars}$ then
  $\s$ $\s$ $\s$ if $\tsc{tailCall}(\term)$
  $\s$ $\s$ $\s$ then $\Susweight$($y$, cont)
  $\s$ $\s$ $\s$ else $\Susweight$($y$,$\lambda x. \tsc{cps}'(\tsf{cont},\term_2)$)
  $\s$ $\s$ else $\ttt{let } x = \term_1 \s \ttt{in} \s \term_2'$$\label{line:cps:endlet}$

function $\tsc{tailCall}$($\term$): $\termanfty \rightarrow \{\false{},\true{}\}$ =
  match $\term$ with
  | $\ttt{let } x = \_ \s \ttt{in} \s x \rightarrow \true{}$
  | $\_ \rightarrow \false{}$
      \end{lstlisting}
    \end{multicols}
  \end{minipage}
\end{algorithm}
\begin{figure}[tb]
  \centering
  \lstset{%
    basicstyle=\ttfamily\scriptsize,
    showlines=true,
    framexleftmargin=-2pt,
    xleftmargin=2em,
  }%
  \begin{multicols}{2}
    \begin{lstlisting}[style=ppl]
let $t_1$ = $2$ in
let $t_2$ = $2$ in
let $t_3$ = $\textrm{Beta}$ in
let $t_4$ = $t_3$ $t_1$ in
let $t_5$ = $t_4$ $t_2$ in
let $a_1$ = assume $t_5$ in
let rec $\mi{iter}$ = $\lambda k.$ $\lambda \mi{obs}.$
  let $t_6$ = $\mi{null}$ in
  let $t_7$ = $t_6$ $\mi{obs}$ in
  if $t_7$ then
    let $t_9$ = $()$ in
    $t_9$
  else
    let $t_{10}$ = $f_\textrm{Bernoulli}$ in
    let $t_{11}$ = $t_{10}$ $a_1$ in
    let $t_{12}$ = $\mi{head}$ in
    let $t_{13}$ = $t_{12}$ $\mi{obs}$ in
    let $t_{14}$ = $t_{11}$ $t_{13}$ in
    $\Susweight($$t_{14}$,
      $\lambda \_$.
        let $t_{15}$ = $\mi{tail}$ in
        let $t_{16}$ = $t_{15}$ $\mi{obs}$ in
        $\mi{iter}$ $k$ $t_{16}$$)$
in
let $t_{18}$ = $\true{}$ in
let $t_{19}$ = $\false{}$ in
let $t_{20}$ = $\true{}$ in
let $t_{21}$ = $\true{}$ in
let $t_{22}$ = $[t_{21}$,$t_{20}$,$t_{19}$,$t_{18}]$ in
let $k'$ = $\lambda \_.$ $a_1$ in
$\mi{iter}$ $k'$ $t_{22}$
    \end{lstlisting}
  \end{multicols}

  \caption{
    The running example from Fig.~\ref{fig:runninganf} after selective CPS transformation.
    The program is semantically equivalent to Fig.~\ref{fig:running:weight}.
  }
  \label{fig:runningcps}
\end{figure}
\noindent
This section presents the second technical contribution: the selective CPS transformation.
The transformations themselves are standard, and the challenge is to correctly use the suspension analysis results for a selective transformation.

Algorithm~\ref{alg:cps} is the full algorithm.
Using terms in ANF as input significantly helps reduce the algorithm's complexity.
The main function \tsc{cps} takes as input a set $\tsf{vars}: \mathcal P(X)$, indicating which expressions to CPS transform, and a program $\term \in \termanfty$ to transform.
It is the new $\tsf{vars}$ argument that separates the transformation from a standard CPS transformation.
For the purposes of this paper, we always use $\tsf{vars} = \{ x \mid \suspend_x = \true{} \}$, where the $\suspend_x$ come from $\tsc{analyzeSuspend}(\term)$.
One could also use $\tsf{vars} = X$ for a standard full CPS transformation (e.g., Fig~\ref{fig:running:full}), or some other set $\tsf{vars}$ for other application domains.
The value returned from the \tsc{cps} function is a (non-ANF) term of the type $T^+$.
The helper function $\tsc{cps}'$, initially called at line~\ref{line:cps:init}, takes as input a continuation term $\tsf{cont}$, indicating the continuation to apply in tail position.
Initially, this continuation term is $\termid$, which indicates no continuation.
Similarly to Algorithm~\ref{alg:gencstr}, the top-level match at line~\ref{line:cps:topmatch} has two cases: a simple case for variables (line~\ref{line:cps:casevar}) and a complex case for \ttt{let} expressions (lines \ref{line:cps:beginlet}--\ref{line:cps:endlet}).
To enable optimization of tail calls, the auxiliary function \tsc{tailCall} indicates whether or not an ANF expression is a tail call (i.e., of the form \ttt{let $x$ = $\term'$ in $x$}).

We now illustrate Algorithm~\ref{alg:cps} by computing $\tsc{cps}(\varsexample, \texample)$, where $\varsexample = \{ \mi{obs}, w_1, t_8, t_{17}, t_{22} \}$ is from~\eqref{eq:analyzerunning}, and $\texample$ is from Fig.~\ref{fig:runninganf}.
Fig.~\ref{fig:runningcps} presents the final result.
First, we note that the transformation does not change expressions not labeled by a name in $\varsexample$, as they do not require suspension.
In the following, we therefore focus only on the transformed expressions.
First, consider the abstraction $\mi{obs}$ defined at line~\ref{line:anflam} in Fig.~\ref{fig:runninganf}, handled by the case at line~\ref{line:cps:abs} in Algorithm~\ref{alg:cps}.
As $\mi{obs} \in \varsexample$, we apply the standard CPS transformation for abstractions: add a continuation parameter to the abstraction and recursively transform the body with this continuation.
Next, consider the transformation of the \ttt{weight} expression $w_1$ at line~\ref{line:weight} in Fig.~\ref{fig:runninganf}, handled by the case at line~\ref{line:cps:weight} in Algorithm~\ref{alg:cps}.
The expression is not at tail position, so we build a new continuation containing the subsequent \ttt{let} expressions, recursively transform the body of the continuation, and then wrap the end result in a Suspension object.
The \ttt{if} expression $t_8$ at line~\ref{line:if} in Fig.~\ref{fig:runninganf}, handled by the case at line~\ref{line:cps:if} in Algorithm~\ref{alg:cps}, is in tail position (it is directly followed by returning $t_8$).
Consequently, we transform both branches recursively.
Finally, we have the applications $t_{17}$ and $t_{22}$ at lines~\ref{line:iter1} and~\ref{line:iter2} in Fig.~\ref{fig:runninganf}, handled by the case at line~\ref{line:cps:app} in Algorithm~\ref{alg:cps}.
The application $t_{17}$ is at tail position, and we transform it by adding the current continuation as an argument.
The application at $t_{22}$ is not at tail position, so we construct a continuation $k'$ that returns the final value $a_1$ (line~\ref{line:a1} in Fig.~\ref{fig:runninganf}), and then add it as an argument to the application.

It is not guaranteed that Algorithm~\ref{alg:cps} produces a correct result.
Specifically, for all applications $\mi{lhs}$ $\mi{rhs}$, we must ensure that (i) if we CPS transform the application, we must also CPS transform \emph{all} possible abstractions that can occur at $\mi{lhs}$, and (ii) if we do \emph{not} CPS transform the application, we must \emph{not} CPS transform any abstraction that can occur at $\mi{lhs}$.
We control this through the argument \tsf{vars}.
In particular, assigning \tsf{vars} according to the suspension analysis produces a correct result.
To see this, consider the application constraints at lines~\ref{line:genappb}--\ref{line:genappe} in Algorithm~\ref{alg:gencstr} again, and note that if any abstraction or intrinsic operation that requires suspension occur at $\mi{lhs}$, $\suspend_x = \true{}$.
Furthermore, the last application constraint ensures that if $\suspend_x = \true{}$, then \emph{all} abstractions and intrinsic operations that occur at $\mi{lhs}$ require suspension.
Consequently, for all $\lambda y. \, \_$ and $const_y \, \_$, either all $\suspend_y = \true{}$ or all $\suspend_y = \false{}$.

\section{Implementation}\label{sec:implementation}
\begin{figure}[tb]
  \centering
  \includegraphics[width=0.9\textwidth]{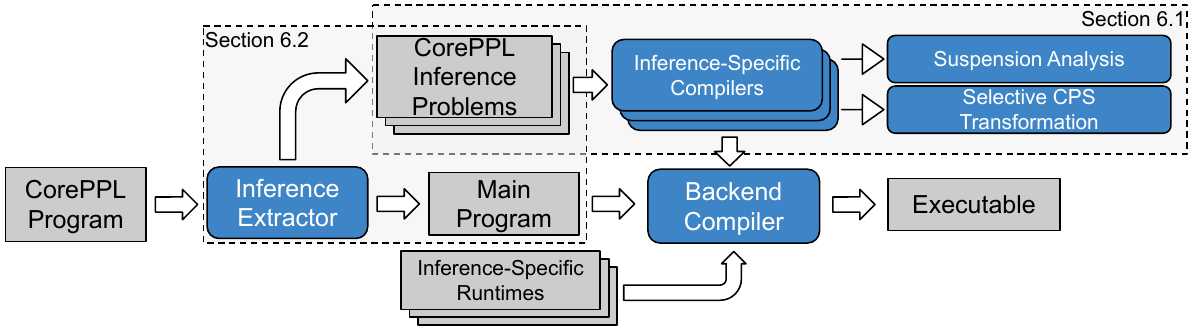}
  \caption{Overview of the Miking CorePPL compiler implementation. We divide the overall compiler into two parts, (i) \emph{suspension analysis and selective CPS} (Section~\ref{sec:susimpl}) and (ii) \emph{inference problem extraction} (Section~\ref{sec:extractimpl}). The figure depicts artifacts as gray rectangular boxes and transformation units and libraries as blue rounded boxes. Note how the \emph{inference extractors} transformation separates the program into two different paths that are combined again after the inference-specific compilation. The white inheritance arrows (pointing to \emph{suspension analysis} and \emph{selective CPS transformations}) mean that these libraries are used within the inference-specific compiler transformation. }
  \label{fig:cppl-overview}
\end{figure}

We implement the suspension analysis and selective CPS transformation in Miking CorePPL~\cite{lunden2022compiling}, a core PPL implemented in the domain-specific language construction framework Miking~\cite{broman2019vision}.
We choose Miking CorePPL for the implementation over other CPS-based PPLs, as the language implementation contains an existing 0-CFA base implementation which simplifies the suspension analysis implementation.
Fig.~\ref{fig:cppl-overview} presents the organization of the CorePPL compiler.
The input is a CorePPL program that may contain many inference problems and applications of inference algorithms, similar to WebPPL and Anglican.
The output is an executable produced by one of the Miking backend compilers.
Section~\ref{sec:susimpl} gives the details of the suspension analysis and selective CPS implementations, and in particular the differences compared to the core calculus in Section~\ref{sec:ppl}.
Section~\ref{sec:extractimpl} presents the inference extractor and its operation combined with selective CPS.
The suspension analysis, selective CPS transformation, and inference extraction implementations consist of roughly 1500 lines of code (a contribution in this paper).
The code is available on GitHub~\cite{mikingdpplgithub}.

\subsection{Suspension Analysis and Selective CPS}\label{sec:susimpl}
Miking CorePPL extends the abstract syntax in Definition~\ref{def:terms} with standard functional data structures and features such as algebraic data types (records, tuples, and variants), lists, and pattern matching.
The suspension analysis and selective CPS implementations in Miking CorePPL extend Algorithm~\ref{alg:gencstr} and Algorithm~\ref{alg:cps} to support these language features.
Furthermore, compared to $\suspend_\ttt{weight}$ and $\suspend_\ttt{assume}$ in Fig.~\ref{fig:semantics}, the implementation allows arbitrary configuration of suspension sources.
In particular, the implementation uses this arbitrary configuration together with the alignment analysis by Lundén et al.~\cite{lunden2023automatic}.
This combination allows selectively CPS transforming to suspend at a subset of \ttt{assume}s or \ttt{weight}s for aligned versions of SMC and MCMC inference algorithms.

Miking CorePPL also includes a framework for inference algorithm implementation.
Specifically, to implement new inference algorithms, users implement an \emph{inference-specific compiler} and \emph{inference-specific runtime}.
Fig.~\ref{fig:cppl-overview} illustrates the different compilers and runtimes.
Each inference-specific compiler applies the suspension analysis and selective CPS transformation to suit the inference algorithm's particular suspension requirements.

Next, we show how Miking CorePPL handles programs containing many inference problems solved with different inference algorithms.

\subsection{Inference Problem Extraction}\label{sec:extractimpl}
Fig.~\ref{fig:cppl-overview} includes the inference extraction compiler procedure.
First, the compiler applies an inference extractor to the input program.
The result is a set of inference problems and a main program containing remaining glue code.
Second, the compiler applies inference-specific compilers to each inference problem.
Finally, the compiler combines the main program and the compiled inference problems with inference-specific runtimes and supplies the result to a backend compiler.

\begin{figure}[tb]
  \centering
  \lstset{%
    language=CorePPL,
    basicstyle=\ttfamily\scriptsize,
    numberstyle=\tiny,
    escapeinside={--}{\^^M},
    showlines=true,
    framexleftmargin=-2pt,
    xleftmargin=2em,
  }%
  \begin{subfigure}{\textwidth}
    \begin{multicols}{2}
\begin{lstlisting}
mexpr
let data = [
  24.0, 42.2, 96.7, 9.2, 85.8,
  34.2, 41.7, 53.4, 85.6, 45.4
] in

let m = lam d. lam y. lam.--\label{fig:extract-ex:m1}
 let x = assume d in
 observe y (Gaussian x 0.1);
 x in--\label{fig:extract-ex:m2}

let d0 =--\label{fig:extract-ex:d1}
 infer (LightweightMCMC--\label{fig:extract-ex:infer1}
  { iterations = 100,
    aligned = true })
  (m (Uniform 0.0 4.0) 1.0) in--\label{fig:extract-ex:d1end}
--\columnbreak
recursive let repeat =
 lam data. lam d.--\label{fig:extract-ex:rep1}
  match data with [y] ++ data then
   let posterior =--\label{fig:extract-ex:d2}
    infer (BPF {particles = 100})--\label{fig:extract-ex:infer2}
           (m d y) in--\label{fig:extract-ex:d2end}
   repeat data posterior--\label{fig:extract-ex:rep2}
  else d
let d1 = repeat data d0 in
match distEmpiricalSamples d1
with (samples, weights) in
iter--\label{fig:extract-ex:dist-print1}
 (lam s.
   print
    (concat (float2string s) "\n"))
 samples--\label{fig:extract-ex:dist-print2}
\end{lstlisting}
    \end{multicols}
    \caption{Miking CorePPL program.}
    \label{fig:extract-ex}
  \end{subfigure}\\
  \begin{subfigure}{.45\textwidth}
\begin{lstlisting}
let m = lam d. lam y. lam.
  let x = assume d in
  observe y (Gaussian x 0.1);
  x in
m (Uniform 0.0 4.0) 1.0 ()
\end{lstlisting}
    \caption{Extracted inference problem from line~\ref{fig:extract-ex:infer1} in (a).}
    \label{fig:extract-ex-infer1}
  \end{subfigure}
  \hspace{7mm}
  \begin{subfigure}{.47\textwidth}
\begin{lstlisting}
let m = lam d. lam y. lam.
  let x = assume d in
  observe y (Gaussian x 0.1);
  x in
m d y ()
\end{lstlisting}
    \caption{Extracted inference problem from line~\ref{fig:extract-ex:infer2} in (a).}
    \label{fig:extract-ex-infer2}
  \end{subfigure}
  \caption{%
    Example Miking CorePPL program in (a) with two non-trivial uses of \ttt{infer}.
    Figures (b) and (c) show the extracted and selectively CPS-transformed inference problems at lines~\ref{fig:extract-ex:infer1} and \ref{fig:extract-ex:infer2} in (a), respectively.
    The compiler handles the free variables \ttt{d} and \ttt{y} in (c) in a later stage.
  }
  \label{fig:extract}
\end{figure}

Consider the example in Fig.~\ref{fig:extract-ex}.
We define a function \ttt{m} that constructs a minimal inference problem on lines~\ref{fig:extract-ex:m1}--\ref{fig:extract-ex:m2}, using a single call to \ttt{assume} and a single call to \ttt{observe} (modifying the execution weight similar to \ttt{weight}).
The function takes an initial probability distribution \ttt{d} and a data point \ttt{y} as input.
We apply aligned lightweight MCMC inference for the inference problem through the \ttt{infer} construct on lines~\ref{fig:extract-ex:d1}--\ref{fig:extract-ex:d1end}.
The first argument to \ttt{infer} gives the inference algorithm configuration, and the second argument the inference problem.
Inference problems are thunks (i.e., functions with a dummy unit argument).
We construct the inference problem thunk by an application of \ttt{m} with a uniform initial distribution and data point $1.0$.
The inference result \ttt{d0} is another probability distribution, and we use it as the first initial distribution in the recursive \ttt{repeat} function (lines~\ref{fig:extract-ex:rep1}--\ref{fig:extract-ex:rep2}).
This function repeatedly performs inference using the SMC bootstrap particle filter (lines~\ref{fig:extract-ex:d2}--\ref{fig:extract-ex:d2end}), again using the function \ttt{m} to construct the sequence of inference problems.
Each \ttt{infer} application uses the result distribution from the previous iteration as the initial distribution and consumes data points from the \ttt{data} sequence.
We extract and print the samples from the final result distribution \ttt{d1} at lines~\ref{fig:extract-ex:dist-print1}--\ref{fig:extract-ex:dist-print2}.
A limitation with the current extraction approach is that we do not yet support nested \ttt{infer}s.

A key challenge in the compiler design is how to handle different inference algorithms within one probabilistic program.
In particular, inference algorithms require different selective CPS transformations, applied to different parts of the code.
To allow the separate handling of inference algorithms, we apply the extraction approach by Hummelgren et al.~\cite{hummelgren2022expression} on the \ttt{infer} applications, producing separate inference problems for each occurrence of \ttt{infer}.
Although the compiler design mostly concerns rather comprehensive engineering work, special care must be taken to handle the non-trivial problem of name bindings when transforming and combining different code entities together.
For instance, the compiler must selectively CPS transform Fig.~\ref{fig:extract-ex-infer1} to suspend at \ttt{assume} (required by MCMC) and selectively CPS transform Fig.~\ref{fig:extract-ex-infer2} to suspend at \ttt{observe} (required by SMC).
We design a robust and modular solution, where it is possible to easily add new inference algorithms without worrying about name conflicts.

\section{Evaluation}\label{sec:evaluation}
This section presents the evaluation of the suspension analysis and selective CPS implementations.
Our main claims are that (i) the approach of selective CPS significantly improves performance compared to traditional full CPS, and (ii) that this holds for a significant set of inference algorithms, evaluated on realistic inference problems.
We use four PPL models and corresponding data sets from the Miking benchmarks repository, available on GitHub~\cite{mikingbenchgithub}.
The models are: constant rate birth-death (CRBD) in Section~\ref{sec:expcrbd}, cladogenetic diversification rate shift (ClaDS) in Section~\ref{sec:expclads}, latent Dirichlet allocation (LDA) in Section~\ref{sec:explda}, and vector-borne disease (VBD) in Section~\ref{sec:expvbd}.
All models are significant and actively used in different research areas:
CRBD and ClaDS in evolutionary biology and phylogenetics~\cite{nee2006birth,ronquist2021universal,maliet2019model}, LDA in topic modeling~\cite{blei2003latent}, and VBD in epidemiology~\cite{funk2016comparative,murray2018delayed}.
In addition to the Miking CorePPL models from the Miking benchmarks, we also implement CRBD in WebPPL and Anglican.

We add a number of popular inference algorithms in Miking CorePPL with support for selective CPS.
The first is standard likelihood weighting (LW), as introduced in Section~\ref{sec:motivating}.
LW does not strictly require CPS, but we implement it with suspensions at \ttt{weight} to highlight the difference between no CPS, selective CPS, and full CPS.
LW gives a good direct measure of CPS overhead as the algorithm simply executes programs many times.
Suspending at \ttt{weight} can also be useful in LW to stop executions with weight 0 (i.e., useless samples) early.
However, we do not use early stopping to isolate the effect CPS has on execution time.
Next, we add the bootstrap particle filter (BPF) and alive particle filter (APF).
Both are SMC algorithms that suspend at \ttt{weight} to resample executions.
BPF is a standard algorithm often used in PPLs, and APF is a related algorithm introduced in a PPL context by Kudlicka et al.~\cite{kudlicka2019probabilistic}.
The final two inference algorithms we add are aligned lightweight MCMC (just MCMC for short) and particle-independent Metropolis--Hastings.
Aligned lightweight MCMC~\cite{lunden2023automatic} is an extension to the standard PPL Metropolis--Hastings approach introduced by Wingate et al.~\cite{wingate2011lightweight}, and suspends at a subset of calls to \ttt{assume}.
Particle-independent Metropolis--Hastings (PIMH) is an MCMC algorithm that repeatedly uses the BPF (suspending at \ttt{weight}) within a Metropolis--Hastings MCMC algorithm~\cite{paige2014compilation}.
We limit the scope to single-core CPU inference.

In addition to the inference algorithms in Miking CorePPL, we also use three other state-of-the-art PPLs for CRBD: Anglican, WebPPL, and the special high-performance RootPPL compiler for Miking CorePPL~\cite{lunden2022compiling}.
For Anglican, we apply LW, BPF, and PIMH inference.
For WebPPL, we use BPF and (non-aligned) lightweight MCMC.
For the RootPPL version of Miking CorePPL, we use BPF inference (the only supported inference algorithm).

We consider two configurations for each model: $1\,000$ and $10\,000$ samples.
An exception is for CRBD and ClaDS, where we adjust APF to use $500$ and $5\,000$ samples to make the inference accuracy comparable to the related BPF.
We run each experiment $300$ times (with one warmup run) and measure execution time (excluding compile time).
To justify the efficiency of the suspension analysis and selective CPS transformation that are part of the compiler, we note here that they, combined, run in only $1$--$5$ ms for all models.

The experiments do \emph{not} compare the performance of different inference algorithms.
To do this, one would also need to consider how accurate the inference results are for a given amount of execution time.
Accuracy varies dramatically between different combinations of inference algorithms and models.
We evaluate the execution time of selective and full CPS in isolation for individual inference algorithms.
Selective CPS is solely an execution time optimization---the algorithms themselves and their accuracy remain unchanged\ifextended\ (we verify this in Appendix~\ref{sec:evalcont} for LW, BPF, and APF).\fi

For Miking CorePPL, we used OCaml 4.12.0 as backend compiler for the implementation in Section~\ref{sec:implementation} and GCC 7.5.0 for the separate RootPPL compiler.
We used Anglican 1.1.0 (OpenJDK 11.0.19) and WebPPL 0.9.15 (Node.js 16.18.0).
We ran the experiments on an Intel Xeon Gold 6148 CPU with 64 GB of memory using Ubuntu 18.04.6.

\subsection{Constant Rate Birth-Death}\label{sec:expcrbd}
\begin{figure}[tb]
  \centering
  \begin{subfigure}{0.47\textwidth}
    \resizebox{\textwidth}{!}{\input{figs/crbd_small_runtime.pgf}}
  \end{subfigure}
  \hspace{5mm}
  \begin{subfigure}{0.47\textwidth}
    \resizebox{\textwidth}{!}{\input{figs/crbd_large_runtime.pgf}}
  \end{subfigure} \\[1mm]
  \renewcommand\s{\hspace{1.5mm}}
  \begin{tabular}{r|cccc}
    & \s Anglican LW \s & \s Anglican BPF \s & \s WebPPL BPF \s & \s WebPPL MCMC \s \\\hline
    $1\,000$ samples & 11.6 $\pm$ 0.36 s & 5.65 $\pm$ 2.71 s & 2.42 $\pm$ 0.20 s & 1.42 $\pm$ 0.07 s \\
    $10\,000$ samples & 90.4 $\pm$ 2.12 s & 29.1 $\pm$ 2.35 s & 53.9 $\pm$ 4.03 s & 3.10 $\pm$ 0.77 s \\
  \end{tabular}
  \caption{%
    Mean execution times for the CRBD model.
    The error bars show 95\% confidence intervals (using the option \ttt{('ci', 95)} in Seaborn's \ttt{barplot}).
    The table shows standard deviations.
  }
  \label{fig:crbd:runtime}
\end{figure}
CRBD is a diversification model, used by evolutionary biologists to infer distributions over birth and death rates for observed evolutionary trees of groups of species, called \emph{phylogenies}.
For the CRBD experiment, we use the Alcedinidae phylogeny (Kingfisher birds, 54 extant species)~\cite{ronquist2021universal,jetz2012global}.
We compare CRBD in Miking CorePPL (55 lines of code), Anglican (129 lines of code), and WebPPL (66 lines of code).
\ifextended The source code is available in Appendix~\ref{sec:expcrbdcont}. \fi
The total experiment execution time was 9 hours.

Fig.~\ref{fig:crbd:runtime} presents the results.
We note that selective CPS is faster than full CPS in all cases.
Unlike full CPS, the overhead of selective CPS compared to no CPS is marginal for LW.
The execution time for early MCMC samples is sensitive to initial conditions, and we therefore see more variance for MCMC compared to the other algorithms.
When we increase the number of samples to $10\,000$, the variance reduces.
With the exception of MCMC in WebPPL, the execution times for Anglican and WebPPL are one order of magnitude slower than the equivalent algorithms in Miking CorePPL.
However, note that the comparison is only for reference and not entirely fair, as Anglican and WebPPL use different execution environments compared to Miking CorePPL.
Lastly, we note that the Miking CorePPL BPF implementation with selective CPS is not much slower than when compiling Miking CorePPL to RootPPL BPF---a compiler designed specifically for efficiency (but with other limitations, such as the lack of garbage collection).
RootPPL does not use CPS, and instead enables suspension through a low-level transformation using the concept of PPL control-flow graphs~\cite{lunden2022compiling}.

\subsection{Cladogenetic Diversification Rate Shift}\label{sec:expclads}
\begin{figure}[tb]
  \begin{subfigure}{0.47\textwidth}
    \resizebox{\textwidth}{!}{\input{figs/clads2_small_runtime.pgf}}
  \end{subfigure}
  \hspace{5mm}
  \begin{subfigure}{0.47\textwidth}
    \resizebox{\textwidth}{!}{\input{figs/clads2_large_runtime.pgf}}
  \end{subfigure}\\
  \vspace{-5mm}
  \caption{%
    Mean execution times for the ClaDS model.
    The error bars show 95\% confidence intervals (using the option \ttt{('ci', 95)} in Seaborn's \ttt{barplot}).
  }
  \label{fig:clads2:runtime}
\end{figure}
ClaDS is another diversification model used in evolutionary biology~\cite{maliet2019model,ronquist2021universal}.
Unlike CRBD, it allows birth and death rates to change over time.
We again use the Alcedinidae phylogeny.
\ifextended
The full source code (72 lines of code) is available in Appendix~\ref{sec:expcladscont}.
\fi
The total experiment execution time was 3 hours.
Fig.~\ref{fig:clads2:runtime} presents the results.
We note that selective CPS is faster than full CPS in all cases.

\subsection{Latent Dirichlet Allocation}\label{sec:explda}
\begin{figure}[tb]
  \begin{subfigure}{0.47\textwidth}
    \resizebox{\textwidth}{!}{\input{figs/lda_small_runtime.pgf}}
  \end{subfigure}
  \hspace{5mm}
  \begin{subfigure}{0.47\textwidth}
    \resizebox{\textwidth}{!}{\input{figs/lda_large_runtime.pgf}}
  \end{subfigure}\\
  \vspace{-5mm}
  \caption{%
    Mean execution times for the LDA model.
    The error bars show 95\% confidence intervals (using the option \ttt{('ci', 95)} in Seaborn's \ttt{barplot}).
  }
  \label{fig:lda:runtime}
\end{figure}

LDA~\cite{blei2003latent} is a model from natural language processing used to categorize documents into \emph{topics}.
We use a synthetic data set with size comparable to the data set in Ritchie et al.~\cite{ritchie2016c3}: a vocabulary of 100 words, 10 topics, and 25 observed documents (30 words in each).
We do not apply any optimization techniques such as collapsed Gibbs sampling~\cite{griffiths2004finding}.
Solving the inference problem using a PPL is therefore challenging already for small data sets.
\ifextended
The full source code (26 lines of code) is available in Appendix~\ref{sec:expldacont}.
\fi
The total experiment execution time was 12 hours.

Fig.~\ref{fig:lda:runtime} presents the results.
We note that selective CPS is faster than full CPS in all cases.
Interestingly, the reduction in overhead compared to full CPS for LW is not as significant.
The reason is that suspension at \ttt{weight} for the model requires that we CPS transform the most computationally expensive recursion.

\subsection{Vector-Borne Disease}\label{sec:expvbd}
\begin{figure}[tb]
  \begin{subfigure}{0.47\textwidth}
    \resizebox{\textwidth}{!}{\input{figs/vbd_small_runtime.pgf}}
  \end{subfigure}
  \hspace{5mm}
  \begin{subfigure}{0.47\textwidth}
    \resizebox{\textwidth}{!}{\input{figs/vbd_large_runtime.pgf}}
  \end{subfigure}\\
  \vspace{-5mm}
  \caption{%
    Mean execution times for the VBD model.
    The error bars show 95\% confidence intervals (using the option \ttt{('ci', 95)} in Seaborn's \ttt{barplot}).
  }
  \label{fig:vbd:runtime}
\end{figure}

We use the VBD model from Funk et al.~\cite{funk2016comparative} and later Murray et al.~\cite{murray2018delayed}.
The background is a dengue outbreak in Micronesia and the spread of disease between mosquitos and humans.
The inference problem is to find the true numbers of susceptible, exposed, infectious, and recovered (SEIR) individuals each day, given daily reported numbers of new cases at health centers.
\ifextended
The full source code (140 lines) is available in Appendix~\ref{sec:expvbdcont}.
\fi
The total execution time was 8 hours.

Fig.~\ref{fig:vbd:runtime} presents the results.
Again, we note that selective CPS is faster than full CPS in all cases, except seemingly for APF and $1\,000$ samples.
This is very likely a statistical anomaly, as the variance for APF is quite severe for the case with $1\,000$ samples.
Compared to the BPF, APF uses a resampling approach for which the execution time varies a lot if the number of samples is too low~\cite{kudlicka2019probabilistic}.
The plots clearly show this as, compared to $1\,000$ samples, the variance is reduced to BPF-comparable levels for $10\,000$ samples.
In summary, the evaluation demonstrates the clear benefits of selective CPS over full CPS for universal PPLs.

\section{Related Work}\label{sec:related}

There are a number of universal PPLs that require non-trivial suspension.
One such language is Anglican~\cite{wood2014new}, which solves the suspension problem using CPS.
Anglican performs a full CPS transformation with one exception---certain statically known functions named \emph{primitive procedures}, that include a subset of the regular Clojure (the host language of Anglican) functions, are guaranteed to not execute PPL code, and Anglican does not CPS transform them~\cite{tolpin2016design}.
However, higher-order functions in Clojure libraries cannot be primitive procedures, and Anglican must manually reimplement such functions (e.g., \ttt{map} and \ttt{fold}).
Anglican does not consider a selective CPS transformation of PPL code, and always fully CPS transforms the PPL part of Anglican programs.

WebPPL~\cite{goodman2014design} and the approach by Ritchie et al.~\cite{ritchie2016c3} also make use of CPS transformations to implement PPL inference.
They do not, however, consider selective CPS transformations.
Ścibior et al.~\cite{scibior2018functional} present an architectural design for a probabilistic functional programming library based on monads and monad transformers (and corresponding theory in Ścibior et al.~\cite{scibior2017denotational}).
In particular, they use a coroutine monad transformer to suspend SMC inference.
This approach is similar to ours in that it makes use of high-level functional language features to enable suspension.
They do not, however, consider a selective transformation.

The PPLs Pyro~\cite{bingham2019pyro}, Stan~\cite{carpenter2017stan,baudart2021compiling}, Gen~\cite{towner2019gen,lew2023smcp3}, and Edward~\cite{tran2016edward} either implement inference algorithms that do not require suspension (e.g., Hamiltonian Monte Carlo), or restrict the language in such a way that suspension is explicit and trivially handled by the language implementation.
For example, SMC in Pyro\footnote{Note that the main inference algorithm in Pyro is stochastic variational inference, which does not require suspension.} and newer versions of Birch require that users explicitly write programs as a \ttt{step} function that the SMC implementation calls iteratively.
Resampling only occurs in between calls to \ttt{step}, and suspension is therefore trivial.

Work on general-purpose selective CPS transformations include Nielsen~\cite{nielsen2001selective}, Asai and Uehara~\cite{asai2017selective}, Rompf et al.~\cite{rompf2009implementing}, and Leijen~\cite{leijen2017type}.
They consider typed languages, unlike the untyped language in this paper.
The early work by Nielsen~\cite{nielsen2001selective} considers the efficient implementation of \ttt{call/cc} through a selective CPS transformation.
The transformation requires manual user annotations, unlike the fully automatic approach in this paper.
A more recent approach is due to Asai and Uehara~\cite{asai2017selective}, who consider an efficient implementation of delimited continuations using \ttt{shift} and \ttt{reset} through a selective CPS transformation.
Similar to us, they automatically determine where to selectively CPS transform programs.
They use an approach based on type inference, while our approach builds upon 0-CFA.
Rompf et al.~\cite{rompf2009implementing} follow a similar approach to Asai and Uehara~\cite{asai2017selective}, but for Scala, and additionally require user annotations.
Leijen~\cite{leijen2017type} uses a type-directed selective CPS transformation to compile algebraic effect handlers.

There are low-level alternatives to CPS for suspension in PPLs.
In particular, there are various languages and approaches that directly implement support for non-preemptive multitasking (e.g., coroutines).
Turing~\cite{ge2018turing} and older versions of Birch~\cite{murray2018automated,murray2020lazy} implement coroutines to enable arbitrary suspension, but do not discuss the implementations in detail.
Lundén et al.~\cite{lunden2022compiling} introduces and uses the concept of PPL control-flow graphs to compile Miking CorePPL to the low-level C++ framework RootPPL.
The compiler explicitly introduces code that maintains special execution call stacks, distinct from the implicit C++ call stacks.
The implementation results in excellent performance, but supports neither garbage collection nor higher-order functions.
Another low-level approach is due to Paige and Wood~\cite{paige2014compilation}, who exploits mutual exclusion locks and the \ttt{fork} system call to suspend and resample SMC executions.
In theory, many of the above low-level alternatives to CPS can, if implemented efficiently, result in the least possible overhead due to more fine-grained low-level control.
The approaches do, however, require significantly more implementation effort compared to a CPS transformation.
Comparatively, the selective CPS transformation is a surprisingly simple, high-level, and easy-to-implement alternative that brings the overhead of CPS closer to that of more low-level approaches.

\section{Conclusion}\label{sec:conclusion}
This paper introduces a selective CPS transformation for the purpose of execution suspension in PPLs.
To enable the transformation, we develop a static suspension analysis that determines parts of programs that require a CPS transformation as a consequence of inference algorithm suspension requirements.
We implement the suspension analysis, selective CPS transformation, and an inference problem extraction procedure (required as a result of the selective CPS transformation) in Miking CorePPL.
Furthermore, we evaluate the implementation on real-world models from phylogenetics, topic-modeling, and epidemiology.
The results demonstrate significant speedups compared to the standard full CPS suspension approach for a large number of Monte Carlo inference algorithms.

\subsubsection{Acknowledgments.}
This project was financially supported by the Swedish Foundation for Strategic Research (FFL15-0032 and RIT15-0012), partially supported by the Swedish Research Council (Grant No. 2018-04329), and by Digital Futures (the DLL project).
The research has also been carried out as part of the Vinnova Competence Center for Trustworthy Edge Computing Systems and Applications at KTH Royal Institute of Technology.
We thank Gizem Çaylak for her LDA implementation and Viktor Senderov for his ClaDS implementation.

\subsubsection{Data-Availability Statement.}
The paper has an accompanying artifact that supports the evaluation: \url{https://zenodo.org/doi/10.5281/zenodo.10454311}.


\bibliographystyle{splncs04}
\clearpage
\bibliography{references}

\ifextended
\clearpage
\appendix

\section{Suspension Analysis, Continued}
This section provides additional details on the suspension analysis.
Specifically, Section~\ref{sec:cfaalg} describes the suspension analysis algorithm, and Section~\ref{sec:proof} presents the proof for Theorem~\ref{thm:main}.

\subsection{Algorithm}\label{sec:cfaalg}
\begin{algorithm}
  \renewcommand{\s}{\hphantom{|}}
  \caption{%
    Suspension analysis.
    We write the functional-style pseudocode for the algorithm itself in sans serif font to distinguish it from terms in $T$.
  }\label{alg:flow}
  \raggedright
\lstinline[style=alg]!function $\tsc{analyzeSuspend}$($\term$): $\termanfty \rightarrow ((X \rightarrow \mathcal P (A)) \times \mathcal P(X))$!\\
  \vspace{-4mm}
  \hspace{3mm}
  \begin{minipage}[t]{0.44\textwidth}
    \begin{lstlisting}[
      name=sanalysis,
      style=alg,
      basicstyle=\sffamily\scriptsize,
      numbers=left,
      showlines=true
    ]
worklist$: [X]$ $\coloneqq$ $[]$
data$: X \rightarrow \mathcal{P}(A)$ $\coloneqq \{(x,\varnothing) \mid x \in X\}$
suspend$: \mathcal{P}(X)$ $\coloneqq \varnothing$
edges$: X \rightarrow \mathcal{P}(R)$ $\coloneqq \{(x,\varnothing) \mid x \in X\}$
for $\cstr \in \tsc{generateConstraints}$($\term$):
  $\tsc{initCstr}(\cstr)$
$\tsc{iter}$(); $\s$ return (data, suspend)

function $\tsc{iter}$: $() \rightarrow ()$ =
  match worklist with
  | $[]$ $\rightarrow ()$
  | $x$ :: worklist' $\rightarrow$
  $\s$ worklist $\coloneqq$ worklist'
  $\s$ for $\cstr$ $\in$ edges(x):
  $\s$ $\tsc{propCstr}$($\cstr$)
  $\s$ $\tsc{iter}$ $()$

function $\tsc{initCstr}$($\cstr$): $R \rightarrow ()$ =
  match $\cstr$ with
  | $\absval \in S_x \rightarrow$ $\tsc{addData}$($x$, $\{\absval\}$)
  | $S_x \subseteq S_y \rightarrow$
  $\s$ $\tsc{initCstr}'$($x$, $\cstr$)
  | $\absval_1 \in S_x \Rightarrow \absval_2 \in S_y \rightarrow$
  $\s$ $\tsc{initCstr}'$($x$, $\cstr$)
  | $\suspend_x \rightarrow$ $\tsc{addSuspend}$($x$)
  | $\suspend_x \Rightarrow \suspend_y \rightarrow$
  $\s$ $\tsc{initCstr}'$($x$, $\cstr$)
  | $\forall x \forall y \s \lambda x. y \in S_\mi{lhs}$
  $\s$ $\s$ $\Rightarrow (S_\mi{rhs} \subseteq S_x) \land (S_y \subseteq S_\mi{app}) \rightarrow$
  $\s$ $\tsc{initCstr}'$($\mi{lhs}$, $\cstr$)
  | $\forall x \forall n \s (\ttt{const}_x n \in S_\mi{lhs}) \land (n > 1)$
  $\s$ $\s$ $\Rightarrow \ttt{const}_x n-1 \in S_\mi{app} \rightarrow$
  $\s$ $\tsc{initCstr}'$($\mi{lhs}$, $\cstr$)
  | $\forall x \s \lambda x. \_ \in S_{\mi{lhs}}$
  $\s$ $\s$ $\Rightarrow (\suspend_x \Rightarrow \suspend_\mi{res}) \rightarrow$
  $\s$ $\tsc{initCstr}'$($\mi{lhs}$, $\cstr$)
  | $\forall x \s \lambda x. \_ \in S_{\mi{lhs}} \Rightarrow \suspend_x \rightarrow$
  $\s$ $\tsc{initCstr}'$($\mi{lhs}$, $\cstr$)
  | $\forall x \s \ttt{const}_x \s \_ \in S_{\mi{lhs}}$
  $\s$ $\s$ $\Rightarrow (\suspend_x \Rightarrow \suspend_\mi{res}) \rightarrow$
  $\s$ $\tsc{initCstr}'$($\mi{lhs}$, $\cstr$)
  | $\forall x \s \ttt{const}_x \s \_ \in S_{\mi{lhs}} \Rightarrow \suspend_x \rightarrow$
  $\s$ $\tsc{initCstr}'$($\mi{lhs}$, $\cstr$)
  | $\suspend_\mi{res} \Rightarrow$
  $\s$ $\s$ $(\forall x \s \lambda x. \_ \in S_{\mi{lhs}} \Rightarrow \suspend_x)$
  $\s$ $\s$ $\land \s (\forall x \s \ttt{const}_x \s \_ \in S_{\mi{lhs}} \Rightarrow \suspend_x)$
  $\s$ $\rightarrow$ $\tsc{initCstr}'$($\mi{res}$, $\cstr$)
    \end{lstlisting}
  \end{minipage}
  \begin{minipage}[t]{0.51\textwidth}
    \begin{lstlisting}[
      name=sanalysis,
      style=alg,
      basicstyle=\sffamily\scriptsize,
      numbers=left,
      showlines=true
    ]
function $\tsc{initCstr}'$($x$,$\cstr$): $X \rightarrow ()$ =
  edges($x$) $\coloneqq$ edges($x$) $\cup \s \{\cstr\}$
  $\tsc{propCstr}$($\cstr$)

function $\tsc{addData}$($x$, A): $X \times \mathcal P(A) \rightarrow ()$ =
  if A $\not \subseteq$ data($x$) then
    data($x$) $\coloneqq$ data($x$) $\cup \s A$
    worklist $\coloneqq$ $x$ $::$ worklist

function $\tsc{addSuspend}$($x$): $X \rightarrow ()$ =
  if $x \not \in$ suspend then
    suspend $\coloneqq$ suspend $\cup \{ x\}$
    worklist $\coloneqq$ $x$ $::$ worklist

function $\tsc{propCstr}$($\cstr$): $R \rightarrow ()$ =
  match $\cstr$ with
  | $\absval \in S_x \rightarrow ()$
  | $S_x \subseteq S_y \rightarrow$ $\tsc{addData}$($y$, data($x$))
  | $\absval_1 \in S_x \Rightarrow \absval_2 \in S_y \rightarrow$
  $\s$ if a$_1 \in$ data($x$) then $\tsc{addData}$($y$,$\{\absval_2\}$)
  | $\suspend_x \rightarrow ()$
  | $\suspend_x \Rightarrow \suspend_y \rightarrow$
  $\s$ if $x \in$ suspend then $\tsc{addSuspend}$($y$)
  | $\forall x \forall y \s \lambda x. y \in S_\mi{lhs} \Rightarrow$
  $\s$ $\s$ $ (S_\mi{rhs} \subseteq S_x) \land (S_y \subseteq S_\mi{app}) \rightarrow$ $\label{line:proplambda}$
  $\s$ for $\lambda x. y \in$ data($\mi{lhs}$):
  $\s$ $\s$ $\tsc{initCstr}$($S_\mi{rhs} \subseteq S_x$)
  $\s$ $\s$ $\tsc{initCstr}$($S_y \subseteq S_\mi{app}$)
  | $\forall x \forall n \s (\ttt{const}_x \, n \in S_\mi{lhs}) \land (n > 1)$
  $\s$ $\s$ $\Rightarrow \ttt{const}_x \, n-1 \in S_\mi{app} \rightarrow$
  $\s$ for $\ttt{const}_x \, n \in$ data($\mi{lhs}$):
  $\s$ $\s$ if $n > 1$ then $\tsc{addData}$($\mi{app}$, $\{\ttt{const}_x \, n-1\}$)
  | $\forall x \s \lambda x. \_ \in S_{\mi{lhs}} \Rightarrow (\suspend_x \Rightarrow \suspend_\mi{res}) \rightarrow$
  $\s$ for $\lambda x. \_ \in$ data($\mi{lhs}$):
  $\s$ $\s$ $\tsc{initCstr}$($\suspend_x \Rightarrow \suspend_\mi{res}$)
  | $\forall x \s \lambda x. \_ \in S_{\mi{lhs}} \Rightarrow \suspend_x \rightarrow$
  $\s$ for $\lambda x. \_ \in$ data($\mi{lhs}$): $\tsc{addSuspend}$($x$)
  | $\forall x \s \ttt{const}_x \s \_ \in S_{\mi{lhs}}$
  $\s$ $\s$ $\Rightarrow (\suspend_x \Rightarrow \suspend_\mi{res}) \rightarrow$
  $\s$ for $\ttt{const}_x \, n \in$ data($\mi{lhs}$):
  $\s$ $\s$ $\tsc{initCstr}$($\suspend_x \Rightarrow \suspend_\mi{res}$)
  | $\forall x \s \ttt{const}_x \s \_ \in S_{\mi{lhs}} \Rightarrow \suspend_x \rightarrow$
  $\s$ for $\ttt{const}_x \, n \in$ data($\mi{lhs}$): $\tsc{addSuspend}$($x$)
  | $\suspend_\mi{res} \Rightarrow (\forall x \s \lambda x. \_ \in S_{\mi{lhs}} \Rightarrow \suspend_x)$
  $\s$ $\s$ $\land (\forall x \s \ttt{const}_x \s \_ \in S_{\mi{lhs}} \Rightarrow \suspend_x) \rightarrow$
  $\s$ if $\mi{res} \in$ suspend then
  $\s$ $\s$ $\tsc{initCstr}$($\forall x \s \lambda x. \_ \in S_{\mi{lhs}} \Rightarrow \suspend_x$)
  $\s$ $\s$ $\tsc{initCstr}$($\forall x \s \ttt{const}_x \s \_ \in S_{\mi{lhs}} \Rightarrow \suspend_x$)
    \end{lstlisting}
  \end{minipage}
\end{algorithm}
\noindent
Before proceeding to the algorithm, we formally define constraints.
\begin{definition}[Constraints]
  We define the constraints $\cstr \in R$ as follows.
  {\upshape
    \begin{equation}\label{eq:cstr}
      \begin{aligned}
        &\begin{aligned}
          \cstr \Coloneqq& \s
          \absval \in S_x
          \s | \s
          S_x \subseteq S_y
          \s | \s
          \absval \in S_x \Rightarrow \absval \in S_y
          \\ |& \s
          \suspend_x
          \s | \s
          \suspend_x \Rightarrow \suspend_y
          \\ |& \s
          \forall x \forall y \s \lambda x. y \in S_{\mi{lhs}} \Rightarrow (S_\mi{rhs} \subseteq S_x) \land (S_y \subseteq S_\mi{app})
          \\ |& \s
          \forall x \forall n \s (\ttt{const}_x \, n \in S_\mi{lhs}) \land (n > 1) \Rightarrow \ttt{const}_x \, n-1 \in S_\mi{app}
          \\ |& \s
          \forall x \s \lambda x. \_ \in S_{\mi{lhs}} \Rightarrow (\suspend_x \Rightarrow \suspend_\mi{res})
          \\ |& \s
          \forall x \s \lambda x. \_ \in S_{\mi{lhs}} \Rightarrow \suspend_x
          \\ |& \s
          \forall x \s \ttt{const}_x \s \_ \in S_{\mi{lhs}} \Rightarrow (\suspend_x \Rightarrow \suspend_\mi{res})
          \\ |& \s
          \forall x \s \ttt{const}_x \s \_ \in S_{\mi{lhs}} \Rightarrow \suspend_x
          \\ |& \s
          \suspend_\mi{res} \Rightarrow
          (\forall x \s \lambda x. \_ \in S_{\mi{lhs}} \Rightarrow \suspend_x)
          \\
          & \hspace{30mm}\land (\forall x \s \ttt{const}_x \s \_ \in S_{\mi{lhs}} \Rightarrow \suspend_x)
        \end{aligned} \\
        &
        x,y,\mi{lhs},\mi{rhs},\mi{app},\mi{res} \in X.
      \end{aligned}
    \end{equation}
  }
\end{definition}%
\noindent
Algorithm~\ref{alg:flow} is the full suspension analysis.
The algorithm uses a worklist and constraints produced by Algorithm~\ref{alg:gencstr} to propagate abstract values throughout the program until fixpoint.
In particular, the algorithm propagates the new suspension-related constraints.

\subsection{Correctness Proof}\label{sec:proof}
Lemma~\ref{lemma:suslemma} directly yields Theorem~\ref{thm:main}.

\newcommand\condenv[1]{\tbf{\upshape(C1#1)}}
\newcommand\ressub[1]{\tbf{\upshape(R1#1)}}
\newcommand\resexists[1]{\tbf{\upshape(R2#1)}}
\newcommand\resret[1]{\tbf{\upshape(R3#1)}}
\newcommand\resall[1]{\ressub{#1}--\resret{#1}}

\begin{lemma}[Suspension analysis soundness]\label{lemma:suslemma}
  Let $\term' \in \termanfty$ be a subterm of $\term$, $\rho \in P$, $s \in S$, $u \in \{\false{},\true{}\}$,$w \in \mathbb{R}$, and $\termv \in V$ such that
  \begin{equation}\label{eq:derivation}
    \rho \vdash \term' \sem{u}{s}{w} \termv.
  \end{equation}
  and for each $x \in X$,
  \begin{description}
    \item[\condenv{}]
      If $\rho(x) = \langle\lambda y. \term_y, \rho_y \rangle$, then $\lambda y. \tsc{name}(\term_y) \in S_x$ and also \condenv{} holds for $\rho_y$.
  \end{description}
  Then,
  \begin{description}
    \item[\ressub{}]
      For every subderivation $(\rho' \vdash \ttt{let } x = \term_1 \ttt{ in } \term_2 \sem{u_1 \lor u_2}{s_1 \concat s_2}{w_1 \cdot w_2} \termv')$ of $(\rho \vdash \term' \sem{u}{s}{w} \termv)$, $u_1 = \true{}$ implies $\suspend_x = \true{}$.
    \item[\resexists{}]
      If $u = \true{}$, then there is an $y \in \tsc{suspendNames}(\term')$ such that $\suspend_y = \true{}$.
    \item[\resret{}]
      If $\termv = \langle\lambda y. \term_y, \rho_y \rangle$, then $\lambda y. \tsc{name}(\term_y) \in S_{\tsc{name}(\term')}$ and also \condenv{} holds for $\rho_y$.
  \end{description}
\end{lemma}
\begin{proof}
  We use structural induction over~\eqref{eq:derivation}.
  First, assume $\term' = x$ and the corresponding derivation
  \[
    \frac{}
    { \rho \vdash x \sem{\false{}}{[]}{1} \rho(x) }
    (\textsc{Var}).
  \]
  Then \ressub{} and \resexists{} holds immediately as there are no subderivations and $u = \false{}$.
  Furthermore, \resret{} holds by \condenv{} and $\tsc{name}(\term') = x$.
  We therefore only need to consider the case $\term' = (\ttt{let } x = \term_1 \ttt{ in } \term_2)$, with derivation
  \newcommand\leftderiv{\rho \vdash \term_1 \sem{u_1}{s_1}{w_1} \termv_1}
  \newcommand\rightderiv{\rho, x \mapsto \termv_1 \vdash \term_2 \sem{u_2}{s_2}{w_2} \termv}
  \newcommand\botderiv{\rho \vdash \ttt{let } x = \term_1 \ttt{ in } \term_2 \sem{u_1 \lor u_2}{s_1 \concat s_2}{w_1 \cdot w_2} \termv}
  \[
    \frac{ \leftderiv \quad \rightderiv} {\botderiv} (\textsc{Let}).
  \]
  To show \ressub{}, we need to show that
  \begin{description}
    \item[\ressub{$_{\term_1}$}] the equivalent of \ressub{} holds for the derivation $\leftderiv$,
    \item[\ressub{$_{\term_2}$}] the equivalent of \ressub{} holds for the derivation $\rightderiv$, and that
    \item[\ressub{$_x$}] $u_1 = \true$ implies $\suspend_x = \true{}$.
  \end{description}
  As $\term_2 \in \termanfty$, we establish \ressub{$_{\term_2}$} by showing the equivalent of \condenv{} for $\rightderiv$, denoted \condenv{$_{\term_2}$}, and applying the induction hypothesis.
  For \resexists{}, consider the case $u_2 = \true$.
  If we establish \condenv{$_{\term_2}$} we get \resexists{$_{\term_2}$} by the induction hypothesis.
  \resexists{} follows as $\tsc{suspendNames}(\term_2) \subseteq \tsc{suspendNames}(\term')$.
  In the following, it is therefore enough to first establish \condenv{$_{\term_2}$} and then assume $u_2 = \false$ when showing \resexists{}.
  Also note that $\resret{}$ follows by $\resret{$_{\term_2}$}$ if \condenv{$_{\term_2}$}, as $\tsc{name}(\term_2) = \tsc{name}(\term')$.
  We now consider each case for $\term_1$.
  To summarize the above, we are done if we establish \ressub{$_{\term_1}$}, \condenv{$_{\term_2}$}, \ressub{$_x$}, and \resexists{} under the assumption $u_2 = \false{}$.
  \newcommand\pcase{\hphantom{l}\\[2mm]\noindent\tbf{Case}}
  \pcase{} $\term_1 = y$\\
  The derivation for $\term_1$ is
  \[
    \frac{}
    { \rho \vdash y \sem{\false{}}{[]}{1} \rho(y) }
    (\textsc{Var})
  \]
  \begin{description}
    \item[\ressub{$_{\term_1}$}] Follows immediately as there are no subderivations.
    \item[\condenv{$_{\term_2}$}] We extend the environment $\rho$ with a binding $x \mapsto \rho(y)$. The result follows by \condenv{} for $\rho$.
    \item[\ressub{$_x$}] Immediate as $u_1 = \false{}$.
    \item[\resexists{}] Immediate as $u_1 = \false{}$.
  \end{description}

  \pcase{} $\term_1 = c$\\
  The derivation for $\term_1$ is
  \[
    \frac{}
    { \rho \vdash c \sem{\false{}}{[]}{1} c }
    (\textsc{Const})
  \]
  \begin{description}
    \item[\ressub{$_{\term_1}$}] Follows immediately as there are no subderivations.
    \item[\condenv{$_{\term_2}$}] We extend the environment $\rho$ with a binding $x \mapsto c$. As $c$ is not an abstraction, the result follows from \condenv{}.
    \item[\ressub{$_x$}] Immediate as $u_1 = \false{}$.
    \item[\resexists{}] Immediate as $u_1 = \false{}$.
  \end{description}

  \pcase{} $\term_1 = \lambda y. \term_y$\\
  The derivation for $\term_1$ is
  \[
    \frac{}
    { \rho \vdash \lambda y. \term_y \sem{\false{}}{[]}{1} \langle\lambda y. \term_y,\rho\rangle }
    (\textsc{Lam})
  \]
  \begin{description}
    \item[\ressub{$_{\term_1}$}] Follows immediately as there are no subderivations.
    \item[\condenv{$_{\term_2}$}]
      We extend the environment $\rho$ with a binding $x \mapsto \langle\lambda y. \term_y,\rho\rangle$.
      By assumption, $\term'$ is a subterm of $\term$, so Lemma~\ref{lemma:cfa} gives $\lambda y. \tsc{name}(\term_y) \in S_x$.
      Furthermore, \condenv{} holds for $rho$ by assumption.
    \item[\ressub{$_x$}] Immediate as $u_1 = \false{}$.
    \item[\resexists{}] Immediate as $u_1 = \false{}$.
  \end{description}
  \pcase{} $\term_1 = y \s z$\\
  The possible derivations for $\term_1$ are
  \newcommand\appbody{\rho_{y'},y' \mapsto \rho(z) \vdash \term_{y'} \sem{u_1}{s_1}{w_1} \termv'}
  \[
    \begin{gathered}
      {\small
        \frac{
          \rho \vdash y \sem{\false{}}{[]}{1} \langle\lambda y'. \term_{y'},\rho_{y'}\rangle \quad
          \rho \vdash z \sem{\false}{[]}{1} \rho(z) \quad
          \appbody
        }
        { \rho \vdash y \s z \sem{u_1}{s_1}{w_1} \termv' }
        (\textsc{App})
      } \\
      \frac{
        \rho \vdash y \sem{\false{}}{[]}{1} c_1 \quad
        \rho \vdash z \sem{\false{}}{[]}{1} c_2
      }
      { \rho \vdash y \s z \sem{\false{}}{[]}{1} \delta(c_1,c_2) }
      (\textsc{Const-App})
    \end{gathered}
  \]
  First, consider the case (\textsc{Const-App}).
  \begin{description}
    \item[\ressub{$_{\term_1}$}] Holds because no subderivation suspends.
    \item[\condenv{$_{\term_2}$}]
      We extend the environment $\rho$ with a binding $x \mapsto \delta(c_1,c_2)$.
      As $\delta(c_1,c_2)$ is not an abstraction, the result follows from \condenv{}.
    \item[\ressub{$_x$}] Immediate as $u_1 = \false{}$.
    \item[\resexists{}] Immediate as $u_1 = \false{}$.
  \end{description}

  Now, consider (\textsc{App}).
  First, note that \condenv{} holds for $\rho_{y'}$ by \condenv{}.
  Also by \condenv{}, $\rho(z)$ fulfills the necessary criteria.
  That is, \condenv{$_{\term_{y'}}$} for the derivation $\appbody$ holds and we apply the induction hypothesis to get \ressub{$_{\term_{y'}}$}, \resexists{$_{\term_{y'}}$}, and \resret{$_{\term_{y'}}$}.
  \begin{description}
    \item[\ressub{$_{\term_1}$}]
      Follows by \ressub{$_{\term_{y'}}$}.
    \item[\condenv{$_{\term_2}$}]
      We extend the environment $\rho$ with a binding $x \mapsto \termv'$.
      By \resret{$_{\term_{y'}}$}, if $\termv' = \langle \lambda y''. \term_{y''}, \rho_{y''}\rangle$, then $\lambda y''. \tsc{name}(\term_{y''}) \in S_{\tsc{name}(\term_{y'})}$.
      Also by \resret{$_{\term_{y'}}$}, \condenv{} holds for $\rho_{y''}$.
      Finally, by Lemma~\ref{lemma:cfa}, we have $S_\tsc{name}{(\term_{y'})} \subseteq S_x$.
      The result follows.
    \item[\ressub{$_x$}]
      Assume $u_1 = \true{}$.
      By \resexists{$_{\term_{y'}}$}, there is an $y'' \in \tsc{suspendNames}(\term_{y'})$ such that $\suspend_y = \true{}$.
      By Lemma~\ref{lemma:cfa},
      $\{ \suspend_n \Rightarrow \suspend_{y'} \mid n \in \tsc{suspendNames}(\term_{y'})\}$.
      As a consequence, $\suspend_{y'} = \true$.
      Furthermore, by \condenv{}, $\lambda y'. \tsc{name}(\term_{y'}) \in S_y$.
      By Lemma~\ref{lemma:cfa}, $\forall y' \, \lambda y'. \_ \in S_{y'} \Rightarrow (\suspend_{y'} \Rightarrow \suspend_x)$.
      It follows that $\suspend_x = \true$, as required.
    \item[\resexists{}]
      As we established earlier, we safely assume $u_2 = \false{}$.
      Now, assume $u_1 = \true{}$.
      The result is immediate by \ressub{$_x$} as $x \in \tsc{suspendNames}(\term')$.
  \end{description}

  \pcase{} $\term_1 = \ttt{if } y \ttt{ then } \term_t \ttt{ else } \term_e$\\
  The possible derivations for $\term_1$ are
  \[
    \begin{gathered}
      \frac{ \rho \vdash y \sem{\false{}}{[]}{1} \true{} \quad \rho \vdash \term_t \sem{u_1}{s_1}{w_1} \termv_t }
      {\rho \vdash \ttt{if } y \ttt{ then } \term_t \ttt{ else } \term_e \sem{u_1}{s_1}{w_1} \termv_t}
      (\textsc{If-True}) \\
      \frac{ \rho \vdash y \sem{\false{}}{[]}{1} \false{} \quad \rho \vdash \term_e \sem{u_1}{s_1}{w_1} \termv_3 }
      {\rho \vdash \ttt{if } y \ttt{ then } \term_t \ttt{ else } \term_e \sem{u_1}{s_1}{w_1} \termv_e}
      (\textsc{If-False})
    \end{gathered}
  \]
  We consider only (\textsc{If-True}) without loss of generality.
  We directly apply the induction hypothesis for $\rho \vdash \term_t \sem{u_1}{s_1}{w_1} \termv_t$ by \condenv{} and get \ressub{$_{\term_t}$}, \resexists{$_{\term_t}$}, and \resret{$_{\term_t}$}.
  \begin{description}
    \item[\ressub{$_{\term_1}$}] Follows by \ressub{$_{\term_t}$}.
    \item[\condenv{$_{\term_2}$}]
      We extend the environment $\rho$ with a binding $x \mapsto \termv_t$.
      By \resret{$_{\term_t}$}, if $\termv_t = \langle \lambda y'. \term_{y'}, \rho_{y'}\rangle$, then $\lambda y'. \tsc{name}(\term_{y'}) \in S_{\tsc{name}(\term_t)}$.
      Also by \resret{$_{\term_t}$}, \condenv{} holds for $\rho_{y'}$.
      Finally, by Lemma~\ref{lemma:cfa}, we have $S_\tsc{name}{(\term_t)} \subseteq S_x$.
      The result follows.
    \item[\ressub{$_x$}]
      Assume $u_1 = \true{}$.
      By \resexists{$_{\term_t}$}, there is an $y' \in \tsc{suspendNames}(\term_t)$ such that $\suspend_y = \true{}$.
      By Lemma~\ref{lemma:cfa},
      $\{ \suspend_n \Rightarrow \suspend_x \mid n \in \tsc{suspendNames}(\term_t)\}$.
      As a consequence, $\suspend_x = \true$.
    \item[\resexists{}]
      Assume $u_1 = \true{}$.
      The result follows as a consequence of \ressub{$_x$} as $x \in \tsc{suspendNames}(\term')$.
  \end{description}
  \pcase{} $\term_1 = \ttt{assume } y$\\
  The derivation for $\term_1$ is
  \[
    \frac{\rho \vdash y \sem{\false{}}{[]}{1} d \quad w' = f_d(c) }
    {\rho \vdash \ttt{assume } y \sem{\suspend_\ttt{assume}}{[c]}{w'} c}
    (\textsc{Assume})
  \]
  \begin{description}
    \item[\ressub{$_{\term_1}$}] Follows immediately as there are no subderivations.
    \item[\condenv{$_{\term_2}$}] We extend the environment $\rho$ with a binding $x \mapsto c$. As $c$ is not an abstraction, the result follows from \condenv{}.
    \item[\ressub{$_x$}] Follows by Lemma~\ref{lemma:cfa}.
    \item[\resexists{}] If $u_1 = \true{}$, then $\suspend_\ttt{assume} = \true{}$ and $x \in \tsc{suspendNames}(\term')$.
  \end{description}
  \pcase{} $\term_1 = \ttt{weight } y$\\
  The derivation for $\term_1$ is
  \[
    \frac{\rho \vdash y \sem{\false{}}{[]}{1} w'}
    {\rho \vdash \ttt{weight } y \sem{\suspend_\ttt{weight}}{s}{w'} ()}
    (\textsc{Weight})
  \]
  \begin{description}
    \item[\ressub{$_{\term_1}$}] Follows immediately as there are no subderivations.
    \item[\condenv{$_{\term_2}$}] We extend the environment $\rho$ with a binding $x \mapsto ()$. As $() \in C$ is not an abstraction, the result follows from \condenv{}.
    \item[\ressub{$_x$}] Follows by Lemma~\ref{lemma:cfa}.
    \item[\resexists{}] If $u_1 = \true{}$, then $\suspend_\ttt{weight} = \true{}$ and $x \in \tsc{suspendNames}(\term')$.
  \end{description}

\end{proof}

\section{Evaluation, Continued}\label{sec:evalcont}
This section presents further details on the evaluation in Section~\ref{sec:evaluation}.
\lstset{%
  showlines=true,
  xleftmargin=1.3em,
}%

\subsection{Constant Rate Birth-Death}\label{sec:expcrbdcont}
Fig.~\ref{fig:crbd_logz} shows violin plot overestimates of the log \emph{marginal likelihood}, also known as the normalizing constant $Z$, for CRBD.
LW, BPF, and APF inference all produce marginal likelihood estimates, and we can use them as a direct measure of inference accuracy and to justify implementation correctness.
We see in Fig.~\ref{fig:crbd_logz} that the distributions of the marginal likelihood estimates are equivalent for each inference algorithm across different PPLs and selective/full CPS, justifying the correctness of the Miking CorePPL implementation.
\begin{figure}[tb]
  \begin{subfigure}{0.47\textwidth}
    \resizebox{\textwidth}{!}{\input{figs/crbd_small_logz.pgf}}
  \end{subfigure}
  \hspace{5mm}
  \begin{subfigure}{0.47\textwidth}
    \resizebox{\textwidth}{!}{\input{figs/crbd_large_logz.pgf}}
  \end{subfigure}
  \caption{Marginal likelihood for the Constant Rate Birth-Death model}
  \label{fig:crbd_logz}
\end{figure}

Listing~\ref{lst:crbdmc}, Listing~\ref{lst:crbdclj}, and Listing~\ref{lst:crbdwppl} give the CRBD source code for Miking CorePPL, Anglican, and WebPPL, respectively.

\begin{lstlisting}[language=CorePPL,basicstyle=\ttfamily\scriptsize,caption=The CorePPL source code for the CRBD experiment in Section~\ref{sec:expcrbd},label=lst:crbdmc]
include "phylo.mc"
include "tree-alcedinidae.mc"
include "math.mc"

mexpr

-- Priors
let lambda = assume (Gamma 1.0 1.0) in
let mu = assume (Gamma 1.0 0.5) in

recursive let survives = lam tBeg.
  let t = subf tBeg (assume (Exponential (addf lambda mu))) in
  if ltf t 0. then
    assume (Bernoulli rho)
  else
    if assume (Bernoulli (divf lambda (addf lambda mu))) then
      if survives t then
        true
      else
        survives t
    else
      false
in

recursive let walk = lam node. lam parentAge.
  let nodeAge = getAge node in
  recursive let simHiddenSpeciation = lam tBeg.
    let t = subf tBeg (assume (Exponential lambda)) in
    if gtf t nodeAge then
      if survives t then
        weight (negf inf)
      else
        weight (log 2.);
        simHiddenSpeciation t
    else ()
  in
  simHiddenSpeciation parentAge;
  observe 0 (Poisson (mulf mu (subf parentAge nodeAge)));
  match node with Node n then
    observe 0. (Exponential lambda);
    resample;
    walk n.left nodeAge;
    walk n.right nodeAge
  else match node with Leaf _ then
    observe true (Bernoulli rho);
    resample
  else never
in

let numLeaves = countLeaves tree in
weight (subf (mulf (subf (int2float numLeaves) 1.) (log 2.))
          (logFactorial numLeaves));
match tree with Node root in
walk root.left root.age;
walk root.right root.age;
lambda
\end{lstlisting}
\begin{lstlisting}[language=Anglican,basicstyle=\ttfamily\scriptsize,caption=The Anglican source code for the CRBD experiment in Section~\ref{sec:expcrbd},label=lst:crbdclj]
(ns crbd.core
  (:require [clojure.tools.cli :refer [parse-opts]])
  (:use [anglican [core :exclude [-main cli-options parse-options]]
         emit runtime]
        clojure.pprint)
  (:gen-class))

(def^:const tree
  ;; Definition omitted for brevity
)

(def^:const rho 0.5684210526315789)

(defdist id
  "Hack to make the factor and condition functions work correctly"
  [] []

  ;; Sampling not allowed
  (sample* [this]
           (throw (Exception. "id-dist does not support sampling")))

  ;; The "log probability" of observing value is value itself
  (observe* [this value] value))

(with-primitive-procedures [id]
  ;; weight
  (defm factor [x]
    "WebPPL-like factor function"
    (observe (id) x))

  (defm condition [b]
    "WebPPL-like condition function"
    (if b (factor 0) (factor Double/NEGATIVE_INFINITY)))

  (defm count-leaves [tree]
    (case (:type tree)
      :node (+ (count-leaves (:left tree)) (count-leaves (:right tree)))
      :leaf 1))

  (defm log-factorial [n] (if (= n 1) 0 (+ (log n) (log-factorial (- n 1))))))

(defquery crbd
  (let [lambda (sample (gamma 1 1))    ;note gamma is parametrized as shape/rate
        mu (sample (gamma 1 2))
        survives (fn survives [t-beg]
                   (let [t (- t-beg (sample (exponential (+ lambda mu))))]
                     (if (< t 0)
                       (sample (flip rho))
                       (if (sample (flip (/ lambda (+ lambda mu))))
                         (or (survives t) (survives t))
                         false))))
        walk (fn walk [tree parent-age]
               (let [sim-hidden-speciation
                     (fn sim-hidden-speciation [t-beg]
                       (let [t (- t-beg (sample (exponential lambda)) )]
                         (if (> t (:age tree))
                           (if (survives t)
                             Double/NEGATIVE_INFINITY
                             (+ (log 2) (sim-hidden-speciation t)))
                           0)))
                     score (+ (sim-hidden-speciation parent-age)
                              (observe*
                               (poisson (* mu (- parent-age (:age tree))))
                               0))]
                 (case (:type tree)
                   :node (do
                           (factor (+ score (observe* (exponential lambda) 0)))
                           (walk (:left tree) (:age tree))
                           (walk (:right tree) (:age tree)))
                   :leaf (factor (+ score (observe* (bernoulli rho) 1))))))
        num-leaves (count-leaves tree)]

    (factor (- (* (- num-leaves 1) (log 2)) (log-factorial num-leaves)))
    (walk (:left tree) (:age tree))
    (walk (:right tree) (:age tree))
    lambda))

(defn norm-const [weights]
  (let [max-weight (apply max weights)]
    (if (= max-weight Double/NEGATIVE_INFINITY)
      max-weight
      (let [sum (reduce + (map #(exp (- % max-weight)) weights))]
        (- (+ max-weight (log sum)) (log (count weights)))))))

(def cli-options
  [["-m" "--method METHOD"
    "Inference method, one of :importance, :pimh, :lmh, or :smc"
    :default :importance
    :parse-fn #(case %
                 ":importance" :importance
                 ":pimh" :pimh
                 ":lmh" :lmh
                 ":smc" :smc)
    :validate [some? "Must be a known algorithm"]]
   ["-p" "--particles COUNT" "Number of particles"
    :default 10
    :parse-fn #(Integer/parseInt %)
    :validate [#(< 0 %) "Must be a number greater than 0"]]
   ["-o" "--output" "Output samples to stdout"]
   ["-h" "--help"]])

(defn int-or-nil [number-string]
  (try (Integer/parseInt number-string) (catch Exception e nil)))

(defn parse-args [args] (if (= (count args) 1) (int-or-nil (first args))))

(defn -main [& args]
  (let [opts (parse-opts args cli-options)
        nsamples (parse-args (:arguments opts))]
    (if (some? (:errors opts))
      (do (println (:errors opts)) (println (:summary opts)) (System/exit 1))
      (if (not (some? nsamples))
        (do (println (:summary opts)) (System/exit 1))
        (let [opts (:options opts)
              samples
              (doall (take nsamples
                           (case (:method opts)
                             (:smc :pimh) (doquery (:method opts)
                                                   crbd
                                                   nil
                                                   :number-of-particles
                                                   (:particles opts)
                                                   :drop-invalid false)
                             (doquery (:method opts) crbd nil
                                      :drop-invalid false))))
              norm-const? (case (:method opts) (:smc :importance) true
                                false)]
          (when norm-const? (println (norm-const (map :log-weight samples))))
          (when (some? (:output opts))
            (run! #(printf "%f %f\n" (:result %) (:log-weight %)) samples)
            (flush)))))))
\end{lstlisting}
\begin{lstlisting}[language=WebPPL,basicstyle=\ttfamily\scriptsize,caption=The WebPPL source code for the CRBD experiment in Section~\ref{sec:expcrbd},label=lst:crbdwppl]
let tree = // Definition omitted for brevity
let rho = 0.5684210526315789

let countLeaves = function(tree) {
    return tree.type == 'node' ?
      countLeaves(tree.left) + countLeaves(tree.right): 1
}

let logFactorial = function(n) {
    return n == 1? 0: Math.log(n) + logFactorial(n - 1)
}

let model = function() {
    // Priors
    let lambda = gamma({shape: 1.0, scale: 1.0})
    let mu = gamma({shape: 1.0, scale: 0.5})

    let survives = function(tBeg) {
        let t = tBeg - exponential({a: lambda + mu})
        if (t < 0) {
            return flip(rho)
        }
        if (flip(lambda/(lambda + mu))) {
            return survives(t) || survives(t)
        }
        return false
    }

    let walk = function(node, parentAge) {
        let simHiddenSpeciation = function(tBeg) {
            let t = tBeg - exponential({a: lambda})
            if (t > node.age) {
                return survives(t)?
                  -Infinity: Math.log(2) + simHiddenSpeciation(t)
            }
            return 0.
        }
        let score = simHiddenSpeciation(parentAge)
                  + Poisson({mu: mu*(parentAge - node.age)}).score(0)
        if (node.type == 'node') {
            factor(score + Exponential({'a': lambda}).score(0))
            walk(node.left, node.age)
            walk(node.right, node.age)
        } else {
            factor(score + Bernoulli({p: rho}).score(true))
        }
    }

    let numLeaves = countLeaves(tree)
    factor((numLeaves - 1)*Math.log(2) - logFactorial(numLeaves))
    walk(tree.left, tree.age)
    walk(tree.right, tree.age)
    return lambda
}

var myArgs = process.argv.slice(3);
var obj = {
  parseInt: parseInt
}
var method = myArgs[0];
var particles = obj.parseInt(myArgs[1], 10);
if (method == 'SMC') {
  var dist = Infer({method: method, particles: particles, model: model})
  dist.normalizationConstant
}
else if (method=='MCMC') {
  var dist = Infer({method: method, samples: particles, model: model})
}
\end{lstlisting}

\subsection{Cladogenetic Diversification Rate Shift}\label{sec:expcladscont}

\begin{figure}[tb]
  \begin{subfigure}{0.47\textwidth}
    \resizebox{\textwidth}{!}{\input{figs/clads2_small_logz.pgf}}
  \end{subfigure}
  \hspace{5mm}
  \begin{subfigure}{0.47\textwidth}
    \resizebox{\textwidth}{!}{\input{figs/clads2_large_logz.pgf}}
  \end{subfigure}
  \caption{Marginal likelihood for the Cladogenetic Diversification Rate Shift model}
  \label{fig:clads2_logz}
\end{figure}
Fig.~\ref{fig:clads2_logz} shows violin plots overestimates of the log marginal likelihood for ClaDS.
We see that the estimates are unchanged across selective and full CPS, justifying the Miking CorePPL implementation.

Listing~\ref{lst:cladsmc} gives the ClaDS source code for Miking CorePPL.

\begin{lstlisting}[language=CorePPL,basicstyle=\ttfamily\scriptsize,caption=The CorePPL source code for the ClaDS experiment in Section~\ref{sec:expclads},label=lst:cladsmc]
include "../../crbd/coreppl/phylo.mc"
include "../../crbd/coreppl/tree-alcedinidae.mc"
include "math.mc"
include "bool.mc"

mexpr

-- Priors
let lambda = assume (Gamma 1.0 1.0) in
let mu = assume (Gamma 1.0 0.5) in
let sigma = sqrt (divf 0.2 (assume (Gamma 1.0 1.0))) in
let log_alpha = assume (Gaussian 0. sigma) in

recursive let survives = lam tBeg. lam multiplier.
  let multiplier =
    mulf multiplier (exp (assume (Gaussian log_alpha sigma))) in
  if or (ltf multiplier 1e-5) (gtf multiplier 1e5) then
    true
  else
    let t =
      subf tBeg
        (assume (Exponential (mulf multiplier (addf lambda mu)))) in
    if ltf t 0. then
      assume (Bernoulli rho)
    else
      if assume (Bernoulli (divf lambda (addf lambda mu))) then
        -- Speciation
        if survives t multiplier then
          true
        else
          survives t multiplier
      else
        -- Extinction
        false
in

recursive let walk = lam node. lam parentAge. lam multiplier.
  let nodeAge = getAge node in
  recursive let simHiddenSpeciation = lam tBeg. lam multiplier.
    let multiplier =
      mulf multiplier (exp (assume (Gaussian log_alpha sigma))) in
    if or (ltf multiplier 1e-5) (gtf multiplier 1e5) then
      weight (negf inf);
      multiplier
    else
      let t = subf tBeg (assume (Exponential (mulf multiplier lambda))) in
      if gtf t nodeAge then
        if survives t multiplier then
          weight (negf inf);
          multiplier
        else
          observe 0 (Poisson (mulf (mulf mu multiplier) (subf tBeg t)));
          weight (log 2.);
          simHiddenSpeciation t multiplier
      else
        observe 0 (Poisson (mulf (mulf mu multiplier) (subf tBeg nodeAge)));
        multiplier
  in
  let multiplier = simHiddenSpeciation parentAge multiplier in
  match node with Node n then
    observe 0. (Exponential (mulf multiplier lambda));
    resample;
    walk n.left nodeAge multiplier;
    walk n.right nodeAge multiplier
  else match node with Leaf _ then
    observe true (Bernoulli rho);
    resample
  else never
in

let numLeaves = countLeaves tree in
weight (subf (mulf (subf (int2float numLeaves) 1.) (log 2.))
         (logFactorial numLeaves));
match tree with Node root in
walk root.left root.age 1.;
walk root.right root.age 1.;
lambda
\end{lstlisting}

\subsection{Latent Dirichlet Allocation}\label{sec:expldacont}

\begin{figure}[tb]
  \begin{subfigure}{0.47\textwidth}
    \resizebox{\textwidth}{!}{\input{figs/lda_small_logz.pgf}}
  \end{subfigure}
  \hspace{5mm}
  \begin{subfigure}{0.47\textwidth}
    \resizebox{\textwidth}{!}{\input{figs/lda_large_logz.pgf}}
  \end{subfigure}
  \caption{Marginal likelihood for the Latent Dirichlet Allocation model}
  \label{fig:lda_logz}
\end{figure}
Fig.~\ref{fig:lda_logz} shows violin plots overestimates of the log marginal likelihood for LDA.
Again, we see that the estimates are unchanged across selective and full CPS, justifying the Miking CorePPL implementation.

Listing~\ref{lst:ldamc} gives the LDA source code for Miking CorePPL.

\begin{lstlisting}[language=CorePPL,basicstyle=\ttfamily\scriptsize,caption=The CorePPL source code for the LDA experiment in Section~\ref{sec:explda},label=lst:ldamc]
include "common.mc"
include "string.mc"
include "seq.mc"
include "ext/dist-ext.mc"

include "data-c3.mc"

mexpr

let alpha: [Float] = make numtopics 1. in
let beta: [Float] = make vocabsize 1. in
let phi = create numtopics (lam. assume (Dirichlet beta)) in
let theta = create numdocs (lam. assume (Dirichlet alpha)) in
repeati (lam w.
    let word = get docs w in
    let counts = assume (Multinomial word.1 (get theta (get docids w))) in
    iteri (lam z. lam e.
        weight (mulf (int2float e)
                  (bernoulliLogPmf (get (get phi z) word.0) true))
      ) counts;
    resample
  ) (length docs);

-- We only compare execution time, and therefore return unit (nothing) as the
-- result
()
\end{lstlisting}

\subsection{Vector-Borne Disease}\label{sec:expvbdcont}

\begin{figure}[tb]
  \begin{subfigure}{0.47\textwidth}
    \resizebox{\textwidth}{!}{\input{figs/vbd_small_logz.pgf}}
  \end{subfigure}
  \hspace{5mm}
  \begin{subfigure}{0.47\textwidth}
    \resizebox{\textwidth}{!}{\input{figs/vbd_large_logz.pgf}}
  \end{subfigure}
  \caption{Marginal likelihood for the Vector-Borne Disease model}
  \label{fig:vbd_logz}
\end{figure}
Fig.~\ref{fig:vbd_logz} shows violin plots overestimates of the log marginal likelihood for VBD.
Again, we see that the estimates are unchanged across selective and full CPS, justifying the Miking CorePPL implementation.

Listing~\ref{lst:vbdmc} gives the VBD source code for Miking CorePPL.

\begin{lstlisting}[language=CorePPL,basicstyle=\ttfamily\scriptsize,caption=The CorePPL source code for the VBD experiment in Section~\ref{sec:expvbd},label=lst:vbdmc]
----------------------------------------------------------------------------
-- The SEIR model from https://docs.birch.sh/examples/VectorBorneDisease/ --
----------------------------------------------------------------------------

-- Include pow and exp
include "math.mc"

-- Include data
include "data.mc"

mexpr

-- Human parameters
let hNu: Float = 0. in
let hMu: Float = 1. in
let hLambda: Float = assume (Beta 1. 1.) in
let hDelta: Float =
  assume (Beta (addf 1. (divf 2. 4.4)) (subf 3. (divf 2. 4.4))) in
let hGamma: Float =
  assume (Beta (addf 1. (divf 2. 4.5)) (subf 3. (divf 2. 4.5))) in

-- Mosquito parameters
let mNu: Float = divf 1. 7. in
let mMu: Float = divf 6. 7. in
let mLambda: Float = assume (Beta 1. 1.) in
let mDelta: Float =
  assume (Beta (addf 1. (divf 2. 6.5)) (subf 3. (divf 2. 6.5))) in
let mGamma: Float = 0. in

-- Other parameters
let rho: Float = assume (Beta 1. 1.) in
let z: Int = 0 in

-- Human SEIR component
let n: Int = 7370 in
let hI: Int = addi 1 (assume (Poisson 5.0)) in
let hE: Int = assume (Poisson 5.0) in
let hR: Int =
  floorfi (assume (Uniform 0. (int2float (addi 1 (subi (subi n hI) hE))))) in
let hS: Int = subi (subi (subi n hE) hI) hR in

-- Human initial deltas
let hDeltaS: Int = 0 in
let hDeltaE: Int = hE in
let hDeltaI: Int = hI in
let hDeltaR: Int = 0 in

-- Mosquito SEIR component
let u: Float = assume (Uniform (negf 1.) 2.) in
let mS: Int = floorfi (mulf (int2float n) (pow 10. u)) in
let mE: Int = 0 in
let mI: Int = 0 in
let mR: Int = 0 in

-- Mosquito initial deltas
let mDeltaS: Int = 0 in
let mDeltaE: Int = 0 in
let mDeltaI: Int = 0 in
let mDeltaR: Int = 0 in

-- Conditioning function
let condition: Int -> Int -> Int -> Int =
  lam t: Int. lam zP: Int. lam hDeltaI: Int.
    let z: Int = addi zP hDeltaI in
    let y: Int = get ys t in
    let z: Int = if neqi (negi 1) y then
      observe y (Binomial z rho); 0
      else z in
    resample;
    z
in

-- Initial conditioning
let z = condition 0 z hDeltaI in

-- Simulation function
recursive let simulate:
  Int -> Int -> Int -> Int -> Int -> Int -> Int -> Int -> Int -> Int -> ()
  =
  lam t: Int.
  lam hSP: Int.
  lam hEP: Int.
  lam hIP: Int.
  lam hRP: Int.
  lam mSP: Int.
  lam mEP: Int.
  lam mIP: Int.
  lam mRP: Int.
  lam zP: Int.

    -- Humans
    let hN: Int = addi (addi (addi hSP hEP) hIP) hRP in
    let hTau: Int =
      assume (Binomial hSP (subf 1. (exp (negf (divf
                                                (int2float mIP)
                                                (int2float hN)))))) in
    let hDeltaE: Int = assume (Binomial hTau hLambda) in
    let hDeltaI: Int = assume (Binomial hEP hDelta) in
    let hDeltaR: Int = assume (Binomial hIP hGamma) in
    let hS: Int = subi hSP hDeltaE in
    let hE: Int = subi (addi hEP hDeltaE) hDeltaI in
    let hI: Int = subi (addi hIP hDeltaI) hDeltaR in
    let hR: Int = addi hRP hDeltaR in

    -- Mosquitos
    let mTau: Int =
      assume (Binomial mSP (subf 1. (exp (negf (divf
                                               (int2float hIP)
                                               (int2float hN)))))) in
    let mN: Int  = addi (addi (addi mSP mEP) mIP) mRP in
    let mDeltaE: Int = assume (Binomial mTau mLambda) in
    let mDeltaI: Int = assume (Binomial mEP mDelta) in
    let mDeltaR: Int = assume (Binomial mIP mGamma) in
    let mS: Int = subi mSP mDeltaE in
    let mE: Int = subi (addi mEP mDeltaE) mDeltaI in
    let mI: Int = subi (addi mIP mDeltaI) mDeltaR in
    let mR: Int = addi mRP mDeltaR in

    let mS: Int = assume (Binomial mS mMu) in
    let mE: Int = assume (Binomial mE mMu) in
    let mI: Int = assume (Binomial mI mMu) in
    let mR: Int = assume (Binomial mR mMu) in

    let mDeltaS: Int = assume (Binomial mN mNu) in
    let mS: Int = addi mS mDeltaS in

    let z: Int = condition t zP hDeltaI in

    -- Recurse
    let tNext: Int = addi t 1 in
    if eqi (length ys) tNext then
      -- We do not return anything here, but simply run the model for
      -- estimating the normalizing constant.
      ()
    else
      simulate tNext hS hE hI hR mS mE mI mR z
in

-- Initiate recursion
simulate 1 hS hE hI hR mS mE mI mR z
\end{lstlisting}

\fi

\end{document}